\documentclass[pdflatex,sn-basic]{sn-jnl}

\usepackage{graphicx}
\usepackage{amsmath,amssymb,amsfonts}
\usepackage{booktabs,tabularx}
\usepackage{textcomp}
\usepackage{placeins}
\usepackage{bookmark}
\usepackage{booktabs}
\usepackage{tabularx}
\usepackage{array}
\usepackage{makecell}

\hypersetup{
  colorlinks=true,
  linkcolor=blue,
  citecolor=blue,
  urlcolor=blue,
  breaklinks=true,
  bookmarksnumbered=true,
  bookmarksopen=true,
  bookmarksopenlevel=2,
  linktoc=all,
  pdfstartview=FitH,
  pdftitle={A Review of Deep-learning-based Seismic Data Denoising and Its Promising Paradigm Shift to Foundation Models},
  pdfauthor={Xintong Dong, Zhengyi Yuan, Changxin Wei, Wenshuo Yu, Shiqi Dong, Jun Lin},
  pdfsubject={Deep-learning-based seismic data denoising and foundation models},
  pdfkeywords={Foundation model, seismic data denoising, pre-training, downstream tasks}
}
\DeclareUnicodeCharacter{FF1A}{:}
\begin{document}

\title[Seismic denoising and foundation models]{A Review of Deep-learning-based Seismic Data Denoising and Its Promising Paradigm Shift to Foundation Models}

\author[1]{\fnm{Xintong} \sur{Dong}}\email{dxt@jlu.edu.cn}
\author[1]{\fnm{Zhengyi} \sur{Yuan}}\email{yuanzhengyi0224@163.com}
\author[1]{\fnm{Changxin} \sur{Wei}}\email{cxwei24@mails.jlu.edu.cn}
\author[1]{\fnm{Wenshuo} \sur{Yu}}\email{yuws24@mails.jlu.edu.cn}
\author*[2]{\fnm{Shiqi} \sur{Dong}}\email{dsq1994@126.com}
\author[1]{\fnm{Jun} \sur{Lin}}\email{lin\_jun@jlu.edu.cn}
\affil[1]{Jilin University, State Key Laboratory of Deep Earth Exploration and Imaging, College of Instrumentation and Electrical Engineering, Changchun, China}
\affil[2]{Northeast Electric Power University, College of Electric Engineering, Jilin, China and Key Laboratory of Modern Power System Simulation and Control and Renewable Energy Technology (Ministry of Education), Jilin, China}
\abstract{Denoising is a long-standing and widely-concerned topic in seismic data processing, for it can significantly increase the signal-to-noise ratio of seismic data. Numerous deep learning (DL) methods have shown promising denoising performance, but most of them are task-specific and focus on a certain type of seismic background noise. The real condition that seismic datasets are often contaminated by various types of noises motivates us to explore a well-generalized and versatile DL model for seismic data denoising. Recently, in the fields of computer vision and natural language processing, foundation models (FMs) pre-trained on vast datasets demonstrate outstanding adaptability and generality across diverse downstream tasks. This paradigm offers a promising path to address the challenges faced by task-specific DL denoising models, such as poor generalization, retraining from scratch for different noise, and the lack of labeled data. We first provide a brief review of traditional seismic denoising methods, followed by a comprehensive review of existing DL-based denoising methods categorized by noise type. Furthermore, we conduct a case study on a dedicated seismic denoising foundation model termed SeisDeFM. This is the first study in geophysical research to develop and validate a seismic denoising foundation model on pre-stack gathers with diverse noise conditions. Experimental results demonstrate that, compared with task-specific DL baselines, SeisDeFM achieves superior denoising performance and cross-noise generalization by the advantages of sufficient pre-training and appropriate downstream adaptation, and it effectively preserves weak reflection events while suppressing complex noise in pre-stack seismic data.}
\maketitle

\bmhead{Article Highlights}
\begin{enumerate}[1.]
\item We review the existing denoising methods and identify the core limitation of task-specific deep learning methods.
\item We present a dedicated seismic denoising foundation model termed SeisDeFM and show a case study on diverse noise attenuation.
\item SeisDeFM achieves superior denoising with small-scale labeled data, enabling efficient large-scale field seismic data processing.
\end{enumerate}

\section{Introduction}\label{sec:introduction}

Seismic data denoising plays a pivotal role in the realm of exploration geophysics. Effective denoising methods can suppress seismic background noise and restore high-fidelity signals that reveal detailed information about subsurface structures and physical parameters. During the acquisition of seismic data, effective signals are simultaneously contaminated by various types of noises that can be roughly divided into random and coherent noises. Coherent noise generated by sources, such as air waves, ground rolls, and multiples, often shows specific propagation characteristics. On the contrary, random noise, primarily caused by environmental interferences, exhibits non-coherency in time or offset.

With the increasing demand for high-precision seismic exploration, the geophysical community has gradually recognized the importance of noise reduction and developed various denoising methods to enhance the quality of seismic data. These seismic denoising methods can be broadly classified into four categories: filtering, decomposition-based, sparse-representation, and low-rank approaches.

\subsection{Traditional seismic denoising methods}\label{sec:traditional-methods}

Filtering methods assume that events of interest are spatially predictable and then use the differences in various domains (time, frequency, offset, and wavenumber, etc) to separate signals from noise. \citet{canalesll1984} introduced the method of random noise attenuation by frequency-space (f-x) predictive filter, showing good performance in several two-dimensional (2D) real examples. \citet{gulunayn1986} further developed the f-x deconvolution based on complex series prediction work proposed by \citet{treitels1974}. \citet{beresfordsmithg1989} applied a frequency-wavenumber (f-k) filter to attenuate the ground rolls, but it is prone to destroy partial signals. \citet{abmar1995} extended the time-space (t-x) and f-x predictive techniques to their three-dimensional (3D) versions. The two 3D extensions enhance noise-attenuation performance by using more samples in the predictions and also relax the strict requirement on linear events. \citet{naghizadehm2012} developed a vector autoregressive model-based f-x denoising method and utilize it to deal with multicomponent seismic data. \citet{cheny2014} proposed a combination of the f-x predictive filter and empirical-mode decomposition (EMD) to enhance the preservation of effective signals while attenuating noise.

EMD \citep{huangne1998} is a classical signal processing technique that can decompose input data into a series of intrinsic modes. This technique specializes in analyzing non-linear and non-stationary signals, which are inherent in seismic data \citep{cheny2014}. \citet{bekaram2009} proposed a data-adaptive EMD-based method for the reduction of random and coherent noise. \citet{hanj2015} applied ensemble EMD to decompose noisy microseismic data and subsequently used adaptive thresholding to remove incoherent components from the resulting intrinsic mode functions. Experimental results showed that this method is effective for denoising both microseismic and reflection seismic records. To accelerate EMD-based denoising implementations, \citet{gomezjl2016} streamlined the computationally expensive polynomial interpolation by introducing a window-averaged sifting operation. Compared to conventional EMD, this improved version reduced computational complexity and required fewer user-defined parameters. Beyond EMD, other decomposition-based methods, such as variational mode decomposition (VMD) and singular value decomposition (SVD), have also been applied to background noise attenuation in seismic data. \citet{bekaram2007} applied local SVD to enhance the signal-to-noise ratio (SNR) of seismic data. Although this method outperformed the f-x deconvolution and median filter for noise suppression, it was less effective in preserving weak and complex events. To overcome the potential energy attenuation of events and low-resolution decomposition of signals when using EMD, \citet{yus2017} developed a novel VMD-based denoising method for random noise. By applying VMD on constant-frequency slices in f-x domain, their approach effectively suppressed random noise while showing improved performance in preserving steep structural slopes.

Sparse-representation approaches assume that seismic data admit a sparse representation in a suitable basis. These methods transform the data into sparsity-promoting domains (e.g., wavelet, Radon, curvelet, contourlet, and shearlet), then remove noise by thresholding the coefficients and applying an inverse transform. \citet{gaoj2006} applied the wavelet transform to denoise the random noise in pre-stack seismic data. Compared with wavelet transform, curvelet transform offers superior directional sensitivity and can capture more geometric features. \citet{hennenfentg2006} extended the fast discrete curvelet transform to non-uniformly sampled data and deployed denoising and interpolation tasks to discuss the sparseness of this extended version on seismic data. Subsequently, \citet{liangx2018}, \citet{zhangc2019}, and \citet{dongx2019} developed several denoising methods based on multi-scale and multi-direction shearlet transform. Dictionary learning adaptively learns basis functions from the data itself, rather than using fixed bases in mathematical transforms, so it can provide a sparser representation for seismic data denoising \citep{dongx2022}. For example, \citet{beckouches2014} partition seismic data into patches and adaptively learn a sparse dictionary, showing superiority over wavelet and curvelet transforms in SNR enhancement and weak signal preservation.

\subsection{DL-based seismic denoising methods}\label{sec:dl-methods}

Most conventional denoising methods rely heavily on manually designed prior assumptions. When these assumptions are violated in real-world applications, these methods often suffer from significant performance degradation \citep{dongx2019,yus2019}. Moreover, tedious empirical parameter tuning further hinders their practical deployment \citep{yus2019,zhongt2024}. Deep learning (DL) methods have shown great promise in the field of artificial general intelligence. Through end-to-end training, they enable the automatic optimization of a massive number of trainable parameters, thereby learning high-dimensional and nonlinear complex mappings \citep{dongx2020a,shengh2025}. Therefore, since 2018, researchers in the relevant fields have developed various DL-based denoising methods for seismic data. In this section, we review these methods according to the types of noise they target.

\subsubsection{Surface waves}\label{sec:surface-waves-review}

Surface waves are a typical type of coherent noise characterized by low velocity and strong energy. Moreover, they often seriously overlap with reflections in both the t-x and f-x domains, which frequently leads to the unintended attenuation of signals during the suppression process. \citet{dongx2019a} and \citet{yus2019} were among the first to employ convolutional neural networks (CNNs) for surface wave attenuation, achieving denoising performance superior to that of conventional approaches. \citet{kaurh2020} generated training labels using a local time-frequency transform and regularized non-stationary regression, and then fed them into a generative adversarial network (GAN). This approach yields results comparable to those of the two conventional label-generation methods, while automating surface wave attenuation and substantially reducing computational cost.\citet{phamn2022} further proposed a physics-constrained DL framework for surface-wave attenuation that integrates unsupervised decomposition, a supervised f-k domain classifier, and signal-to-noise mapping, thereby effectively suppressing noise while preserving weak reflections. \citet{yangl2023b} proposed a fully convolutional framework that leverages dense and skip connections to improve information propagation, thereby better preserving low-frequency signals overlapping with surface waves. \citet{liy2024} proposed a conditional diffusion model that jointly generates clean data and surface waves, so as to minimize reflection leakage and outperforming two conventional methods.

\subsubsection{Random noise}\label{sec:random-noise-review}

Random noise, another major and ubiquitous type, is a spatio-temporal stochastic process arising from natural or anthropogenic sources \citep{lig2017}. The attenuation of random noise remains a challenging task due to its complex characteristics, including the absence of fixed frequencies and apparent velocities, non-stationarity, and non-Gaussianity \citep{zhongt2015}.

In the early stages, experts commonly employed CNN-based architectures and trained them in a supervised fashion. For example, \citet{wangy2019} applied data augmentation to generate paired datasets for network training, achieving better denoising performance than conventional methods. The U-Net, proposed by \citet{ronnebergero2015}, is a widely used encoder-decoder CNN that employs skip connections to preserve fine-grained spatial information. \citet{zhongt2022} later introduced an enhanced version that incorporates residual learning and reconstruction blocks, which outperformed the standard U-Net in both noise attenuation and signal preservation. Afterwards, \citet{zhaoh2023} proposed another improved U-Net by incorporating deformable and dilated convolutions, which outperformed two classic DL-based methods in terms of quantitative analyses and visual quality.

Despite their promising performance, these methods are grounded in supervised learning paradigms that demand a substantial number of high-quality labeled datasets, thereby imposing significant hurdles in dataset sourcing and preparation \citep{sunh2022,zhangy2023}. A number of unsupervised or self-supervised learning approaches have recently been proposed to circumvent this limitation. \citet{birniec2021} introduced a self-supervised blind-spot network that circumvents the need for noisy-clean data pairs. They generated corrupted samples by randomly masking a set of non-adjacent pixels and used the original data as training targets. By recovering these masked pixels from their neighboring ones, this network learns directly from the noisy data without labels, effectively suppressing random noise while minimally distorting the signals. \citet{sunh2022} achieved random noise suppression via a patch-wise self-supervised approach that integrates invariant function and transfer learning, requiring only a single noisy shot gather. \citet{zhangy2023} proposed an unsupervised recursive deep image prior (RDIP) approach that incorporates an improved quality control criterion for random noise attenuation. The RDIP iteratively retrains the network by using the previous noisy input and its corresponding denoised output as a new input-target pair. This recursive self-training strategy progressively refines the result until convergence, yielding higher denoising accuracy and reduced signal leakage.

These aforementioned methods are built upon CNN architectures and are therefore inherently constrained by the local receptive fields of convolutional kernels. To overcome this limitation, more advanced frameworks have been adopted for seismic random noise attenuation. Transformer \citep{vaswania2017}, with self-attention mechanism, can capture more global features. Therefore, \citet{lif2024} adopted Swin-Transformer \citep{liuz2021} to suppress random noise and generated higher SNR than competing methods. \citet{zhangy2024} combined Swin-Transformer with GAN to simultaneously interpolate and denoise seismic data. However, Transformer-based architectures often suffer from quadratic computational complexity, incurring substantial time and computational costs during both training and testing. Recently, a novel architecture named Mamba \citep{gua2024} can model global dependencies with linear complexity, significantly accelerating the training process. \citet{chenh2025} combined the Mamba architecture with fast Fourier convolutions to suppress seismic random noise in both time and frequency domains. This approach achieved superior denoising performance compared to Transformer-based methods, while incurring substantially lower computational cost.

\subsubsection{Erratic noise}\label{sec:erratic-noise-review}

Erratic noise refers to a highly non-Gaussian component of seismic background noise, commonly originating from natural sources (e.g., air blasts, wind, rain, and poor surface conditions), cultural interferences (e.g., power-line noise), acquisition glitches (e.g., recording or parity errors), and quality-control artifacts \citep{tricketts2012}. It usually appears as localized high-amplitude anomalies in seismic records, with its noise level varying from traces, time windows and shot gathers \citep{wangs2022}.

In recent years, geophysical scholars have made considerable efforts to develop DL-based methods for erratic noise suppression. \citet{wangs2022} attenuated erratic noise by introducing an attention-based CNN together with a shuffled noise training data generation strategy, allowing the network to better focus on noisy regions while eliminating the need for clean training data. A joint-guided denoising network was proposed by \citet{zhongt2024} to attenuate erratic noise by enriching feature interactions in CNNs. This network shows good denoising performance under extremely low SNR conditions.

To increase the flexibility of DL-based erratic noise attenuation without the need for paired noisy and clean training data, \citet{qianf2022,qianf2024} turned to unsupervised learning strategies. In their earlier work \citep{qianf2022}, they constructed a robust unsupervised deep convolutional auto-encoder capable of removing both erratic and Gaussian noise. Subsequently, they extended their approach to 3D version by proposing a robust tensor deep learning framework that integrates tensor sparse representation with a tensor neural network, thereby effectively exploiting the 3D spatial correlations of seismic data. However, most of unsupervised denoising methods are prone to converging to sub-optimal solution and often suffer from high computational cost. To address this issue, \citet{baoq2025} further designed a lightweight unsupervised framework based on implicit neural representation. This framework incorporated positional encoding and specialized activation functions to accelerate the procedure of network optimization.

Most of the aforementioned erratic noise attenuation approaches are purely data-driven, with no explicit incorporation of physical constraints. Such a data-centric paradigm inevitably compromises signal fidelity, often leading to waveform distortion and degraded robustness. As a result, whether the reconstructed signals truly honor the underlying physics remains an open question. To alleviate this, \citet{xiep2025} designed a physics-informed neural network to simultaneously remove random and erratic noises. The effective recovery of high-frequency signals on both synthetic and field data highlights the contribution of physical information.

\subsubsection{Multiples}\label{sec:multiples-review}

In field seismic data, the recorded events include both primaries and multiples. Primaries are seismic waves that have undergone exactly one reflection from subsurface interfaces, whereas multiples result from two or more reflections before reaching the receiver \citep{verschuurd2013,durallr2024}. Multiples do not correspond to true subsurface interfaces and are generally eliminated prior imaging and interpretation \citep{liz2021,taol2022,wangk2022}.

Depending on their propagation paths, multiples are broadly classified into surface-related and internal multiples \citep{wegleinab1999}. Most DL-based demultiple methods have focused on the former. Among these, U-Net-based approaches constitute a major category. \citet{liz2021} were among the first to apply U-Net to surface-related multiple removal, achieving a substantial improvement in SNR over conventional linear regression methods. \citet{durallr2024} employed a U-Net to learn the move-out discrimination between primaries and multiples, rather than directly predicting the multiples themselves. Tests on synthetic and field datasets demonstrate that this method effectively suppresses multiples while preserving the amplitude of primaries. The physics-guided DL framework (PGDL) proposed by \citet{zhangd2026} is another U-Net-based method, which integrates the original recorded full wavefield and wave-equation-based multiple predictions as a dual-channel input. PGDL overcomes the limitation that most neural networks typically treat seismic records merely as images, disregarding the physical propagation of seismic waves.

To reduce the dependence on labeled true primaries or multiples, researchers have explored various unsupervised and self-supervised methods. \citet{wangk2022a} proposed an unsupervised deep neural network based on ensemble learning (UDNNEL), which integrates a residual network, a simple U-Net, and an attention-based network. The UDNNEL jointly optimizes the three networks by integrating their six generated outputs, allowing the model to aggregate the advantages of each learner. \citet{liz2023} introduced an unsupervised fast iterative shrinkage thresholding algorithm network (FISTA-Net) for surface-related multiple removal, where the FISTA iterations are unfolded into a neural network and the shrinkage step is replaced by a U-Net. Subsequently, \citet{qiz2025} enhanced FISTA-Net by adopting a semi-supervised strategy that leverages both limited labeled and pseudo-labeled data, achieving markedly better performance than the original unsupervised version.

Several studies have also addressed the more challenging internal multiples. For example, \citet{liux2022} and \citet{wangk2023} proposed U-Net-based supervised and unsupervised methods  for internal multiple suppression, respectively. To further enhance the suppression performance, \citet{zhangm2025} incorporated U-Net with Transformer to fuse both local and global features, yielding better performance and efficiency than existing conventional and CNN-based methods.

Despite the success of these demultiple methods, some researchers have adopted a different perspective, treating multiples as useful signals and exploiting them together with primaries for joint imaging, thus obviating the need for attenuation \citep{verschuurd2016,lus2021}. \citet{youj2025} developed Deepimaging, a DL-based multiple imaging framework. This framework leverages stochastic velocity models as the training foundation, introduces a novel DL-driven approach to represent one-way wave propagation operators, and ultimately establishes an intelligent full-wavefield imaging system.

\subsubsection{DAS denoising}\label{sec:das-review}

Over the past decade, distributed optical-fiber acoustic sensing (DAS) has been proven as a highly promising technology for seismic data acquisition. In contrast to conventional electronic geophones, DAS leverages optical fibers as sensing elements to measure strain induced by seismic waves, offering both high spatial and temporal resolution \citep{nesterovao2026,zhangh2026}. In seismic exploration, DAS has gained acceptance as an alternative to conventional downhole geophones, owing to its exceptional resilience to extreme temperature and pressure. This enables the acquisition of high-fidelity and densely-sampled VSP data with full-well coverage \citep{lelloucha2020,yangl2023a}.

In DAS-VSP data, in addition to conventional seismic background noise, signals are also contaminated by multiple instrument-related noises that are inherent to fiber-optic sensing systems. These instrument-related noises, including fading, horizontal, and checkerboard noises \citep{liy2022}, are absent from other seismic datasets, and their underlying mechanisms remain poorly understood. At that time, no effective suppression techniques were available, creating a critical bottleneck that severely hindered the widespread deployment of DAS acquisition technology.

One of the earliest DL-based DAS-VSP denoising studies was by \citet{dongx2020a}, who proposed an end-to-end CNN with a novel mean square error loss that incorporates an energy ratio matrix. While this method effectively suppresses both random and coherent noises, signal leakage observed in the residual records indicates that signal preservation still requires further improvement. To address this issue, \citet{dongx2020} later developed a convolutional adversarial denoising network (CADN) based on game theory. In CADN, a generator and a discriminator engage in an adversarial process that drives the reconstructed wavefield's probability distribution to asymptotically approach the target, thereby achieving more accurate signal recovery.

The attention mechanism, a widely-used tool in neural networks, dynamically assigns weights to feature maps to focus on the most relevant information during feature extraction \citep{bahdanaud2014}. For DAS-VSP denoising, this mechanism is particularly beneficial, as it can prioritize informative signals over various noise patterns, proving especially effective for recovering weak signals. \citet{liy2022} proposed an attention-based progressive multi-stage denoising network for DAS-VSP data. These cascaded stages progressively refine the denoised results, while the channel and stage-interactive attention mechanisms enhance the feature representation of this network. To handle complex DAS noise, \citet{yangl2023a} integrated dense connections and a kernel-wise attention mechanism into a multi-scale network. Ablation studies validated that the attention module significantly enhances multi-scale feature fusion and consequently boosts denoising performance.

Beyond attention mechanisms, several studies have also turned to unsupervised or self-supervised frameworks that impose fewer constraints on labeled data, thus offering greater flexibility in data preparation. \citet{yangl2023} pioneered an unsupervised denoising network with a U-Net-like architecture, which outperforms both conventional and DL-based counterparts. Subsequently, \citet{saadom2024} developed an unsupervised DL model for DAS-VSP denoising, using band-pass filtered data and continuous wavelet transform (CWT) coefficients as auxiliary guidance. Building on this strategy, \citet{saadom2026} further introduced a self-supervised conditional diffusion model, in which the coarsest-scale CWT serves as conditioning information to steer the reverse diffusion process. Notably, the two approaches \citep{saadom2024,saadom2026} exploit CWT coefficients as an effective prior for signal preservation, and their practical feasibility has been demonstrated through extensive evaluations on field datasets.

The above studies have mostly targeted single-task denoising improvement, but recent efforts have begun exploring integrated frameworks that address multiple DAS-VSP processing tasks jointly. Notably, \citet{chengm2025} proposed a multi-task preprocessing model (MTPM) that unifies denoising, interpolation, and wavefield separation into a single CNN-Transformer network with a two-stage training regimen. Unlike conventional step-by-step pipelines, this approach performs the three tasks concurrently via one-step prediction, effectively preventing cumulative signal attenuation and yielding enhanced overall processing performance.

\subsection{Foundation model for Pre-stack Seismic Data Denoising}\label{sec:foundation-model-review}

Although existing DL-based seismic denoising methods perform well on specific noise types, most are designed for a single noise scenario and fail to generalize to mixed-noise environments. In field seismic exploration, signals are often simultaneously contaminated by multiple types of noises \citep{hlebnikovv2021,chengs2024}. Moreover, even the same type of noise can exhibit significant distribution discrepancies across different survey areas, making it difficult for task-specific models to be directly applied to new areas. Therefore, it is imperative to develop a seismic denoising foundation model (FM) capable of handling diverse noise conditions across different survey areas.

In recent years, FMs have emerged as an important research direction in the field of artificial intelligence. Unlike task-specific DL models, FMs are usually pre-trained on large-scale datasets to learn more general data representations, and then fine-tuned with a small amount of labeled data or task-specific data for different downstream tasks \citep{awaism2025}. Unlike task-specific models, FMs offer superior feature extraction and stronger generalization. In addition, FMs generally have the capability of modeling long-range dependencies \citep{huangy2023}, allowing them to capture global features of different modalities.

Recent advances in FMs have been largely driven by the rapid development of pre-training techniques in natural language processing (NLP) and computer vision (CV). \citet{devlinj2019} proposed a language representation model, bidirectional encoder representations from Transformers (BERT), which learns contextualized representations by pre-training on large-scale text data and achieves excellent performance across various downstream tasks. Subsequently, \citet{brownt2020} proposed a large-scale language model pre-trained on massive text data, namely GPT-3, which demonstrated excellent language understanding capabilities. With only a few examples, GPT-3 can adapt to different NLP tasks. Inspired by the success of Transformer-based architectures in NLP, \citet{dosovitskiya2020} proposed Vision Transformer (ViT), which extended the Transformer architecture to CV tasks. \citet{caronm2021} further introduced the self-supervised DINO framework, enabling visual representation learning from unlabeled images and demonstrating strong generalization capability in object recognition tasks. \citet{kirillova2023} proposed the segment anything model (SAM), which was trained on large-scale image datasets with segmentation annotations and achieved zero-shot segmentation across diverse objects.

However, due to the scarcity of large-scale labeled seismic data, the prohibitive computational cost of pre-training, and the inherent complexity of seismic wavefields, research on foundation models (FMs) for seismic exploration remains in its infancy. Only a limited number of studies have explored the potential of FMs for seismic data processing and interpretation tasks. \citet{harsukor2022} proposed the StorSeismic model, which is based on the BERT \citep{devlinj2019}. The model was pre-trained on pre-stack seismic data and subsequently adapted to downstream tasks by using fine-tuning. \citet{shengh2025} developed seismic FM based on the masked autoencoder architecture \citep{hek2022} and transferred the pre-trained model to various seismic downstream tasks. \citet{dongx2026} proposed a lightweight seismic processing FM (SPFM) based on the Mamba model \citep{gua2024}. SPFM can effectively capture global features of seismic data while alleviating the quadratic computational complexity inherent in Transformer-based models. \citet{chengs2025} introduced a generative seismic FM (GSFM) based on diffusion models. Through probabilistic modeling, GSFM effectively addresses multi-task seismic processing problems and achieves uncertainty quantification.

In this paper, we propose for the first time a new seismic denoising paradigm based on the seismic denoising foundation model (SeisDeFM). This paradigm aims to train the denoising model with massive clean seismic data to learn the intrinsic representation of effective seismic signals, and enable rapid transfer and adaptation to downstream denoising tasks targeting diverse noise types using only a small amount of paired data, thereby meeting the demand for efficient, high-precision processing of industrial-scale massive seismic data. Furthermore, through a case study of SeisDeFM, we comprehensively present and demonstrate its viable architecture, pre-training strategy, downstream task adaptation methods, and dataset construction workflow. The experimental results in the case study verify that the SeisDeFM outperforms task-specific denoising models (TSDM) in training cost, denoising performance, and generalization capability, offering a new paradigm worthy of industry reference and promotion for industrial-scale seismic data denoising.

\section{Principles of SeisDeFM}\label{sec:principles}

In this section, we first introduce the basic theory of SeisDeFM. Subsequently, we elaborate on the framework of SPFM, including training data preparation, viable model architecture, pretraining strategies, downstream adaptation methods, and evaluation metrics for denoising performance.

\subsection{The basic theory of SeisDeFM}\label{sec:basic-theory}

The observed pre-stack seismic data $y\in\mathbb{R}$ can be expressed as the sum of clean signals $x\in\mathbb{R}$ and noises $n\in\mathbb{R}$:

\begin{equation}
y(h)=x+n(h),\quad \forall h\in\{1,2,\dots,H\},
\label{eq:1}
\end{equation}

where $ h $ represents the type of noise. Each type of noise obeys its unique probability distribution:

\begin{equation}
n(h)\sim P_h(\phi_h),
\label{eq:2}
\end{equation}

where$ P_h $represents the specific probability distribution followed by the $ h^{\mathrm{th}}$ type of noises, and $\phi_h$ stands for the set of statistical parameters governing the distribution of noises, including the mean value, variance and higher-order statistics, etc. TSDM is optimized under the empirical risk minimization (ERM) framework, where the optimal model parameters are estimated through neural network training:

\begin{equation}
\hat{\theta}=\arg\min_{\theta}\mathbb{E}_{(y,x)\sim P_{TS}}\left[\mathcal{L}\left(\mathcal{F}_{\theta}\left(y(h)\right),x\right)\right],
\label{eq:3}
\end{equation}

where $\theta$ and $\hat{\theta}$ represents the trainable and optimal parameters of TSDM, respectively, $P_{TS}$ represents the data distribution of the observed pre-stack seismic data and clean signals used for training TSDM, $\mathcal{L}$ represents the loss function, and $\mathcal{F}_{\theta}(\cdot)$ represents the nonlinear mapping of TSDM. Sufficient quantity, feature-complete and labeled training datasets are required to support the training of TSDM for searching the $\hat{\theta}$:

\begin{equation}
\mathcal{D}_{TS}=\left\{\left(y_i(h),x_i\right)\right\}_{i=1}^{N},
\label{eq:4}
\end{equation}

where $\mathcal{D}_{TS}$ represents the dataset used for training TSDM, and $N $ denotes the total number of paired training samples. The intrinsic limitation of TSDM lies in its implicit assumption that $\mathcal{D}_{TS}$ and the test data for field applications are independent and identically-distributed. Given the massive volume of modern seismic data, the diversity of clean data, the complexity of noise, and the scarcity of paired noisy-clean seismic data, the distributions of field data may deviate from pre-defined training datasets, which causes TSDM to underfit field data and to exhibit severe degradation of generalization performance across different surveys and targets.

By employing massive parameterization, large-scale pre-training and small-scale transfer learning, SeisDeFM can break through the capacity bottlenecks of TSDM. It learns physically consistent wavefield representations and rapidly adapts to diverse noise, thereby enabling robust generalization across highly complex and unseen field conditions. Accordingly, SeisDeFM requires abundant training data to learn rich features from diverse field data. The distribution of datasets for training SeisDeFM is denoted as $ P_{FM}$, which can be expressed as:

\begin{equation}
P_{FM}\equiv\sum_{j=1}^{J}\alpha_j P_j,\quad \text{with}\quad \sum_{j=1}^{J}\alpha_j\equiv 1,
\label{eq:5}
\end{equation}

where $ P_j $ represents the data distribution of the $ j^{\mathrm{th}}$ type of dataset and $\alpha_j $ is the weighting coefficient. Besides datasets that contain various types of noise, SeisDeFM requires data categories covering diverse features of clean signals, which yields $ J>H $. Subsequently, SeisDeFM is trained by solving the following optimization problem:

\begin{equation}
\hat{\Theta}=\arg\min_{\Theta}\mathbb{E}_{(y,x)\sim P_{FM}}\left[\mathcal{L}\left(\mathcal{F}_{\Theta}(y),x\right)\right],
\label{eq:6}
\end{equation}

where$\Theta $ and$\hat{\Theta}$ represents the trainable and optimal parameters of SeisDeFM, respectively, and $\mathcal{F}_{\Theta}(\cdot)$ represents the the nonlinear mapping of SeisDeFM. Compared with Equation~\eqref{eq:3}, Equation~\eqref{eq:6} enables SeisDeFM to capture seismic representations that encode local correlations and global dependencies across different data distributions, providing stronger generalization and transfer ability.

The construction and application of SeisDeFM generally follow a two-stage paradigm of pre-training and downstream task adaptation \citep{liuq2025} . This paradigm aims to realize smooth transfer from universal feature representations to task-specific seismic denoising for seismic exploration. To build robust wavefield representations, SeisDeFM is first pre-trained on massive clean seismic data. Such data are either generated by wave-equation-based forward modeling, or obtained from field data, including nearly clean raw data and denoised data processed by conventional methods. In the pre-training stage, SeisDeFM is optimized with a self-supervised loss function:

\begin{equation}
\mathcal{L}_{pre}=\mathbb{E}_{x\sim P_{FM}}\left[\left\|\mathcal{F}_{\Theta}\big(S(x)\big)-x\right\|_{2}^{2}\right],
\label{eq:7}
\end{equation}

where $\mathcal{L}_{pre}$ represents the loss function for SeisDeFM pre-training, $\|\cdot\|_2 $ represents the L2 norm, and $ S(\cdot)$ denotes the operation to encourage SeisDeFM to learn robust wavefield representations, such as masked reconstruction and contrastive learning, etc. During pre-training, SeisDeFM captures seismic features such as event coherency, reflection patterns, and time-frequency correlations. These features are independent of specific noise types. By encoding physical priors of clean wavefields through self-supervised pre-training, SeisDeFM bypasses the need to learn seismic characteristics from scratch, requiring only small-scale paired noisy-clean data to learn specific noise distributions for rapid downstream adaptation. The paired dataset adopted to adapt SeisDeFM to specific downstream denoising tasks for transfer learning can be denoted as:

\begin{equation}
\mathcal{D}_{TS}=\left\{\big(y_i(h),x_i\big)\right\}_{i=1}^{M},\quad M\ll N,
\label{eq:8}
\end{equation}

where $\mathcal{D}_{TS}\sim P_{FM}$. The transfer learning of SeisDeFM is optimized by a supervised loss function:

\begin{equation}
\mathcal{L}_{tl}\equiv \frac{1}{H}\sum_{h=1}^{H}\Bigg\{\mathbb{E}_{(y,x)\sim P_{TS}}\left[\left\|\mathcal{F}_{\Theta}\big(y(h)\big)-x\right\|_{2}^{2}+\lambda\mathcal{L}_{reg}\right]\Bigg\},
\label{eq:9}
\end{equation}

where $\mathcal{L}_{reg}$ and $\lambda$ represent the regularization term adapted to specific signal-to-noise characteristics and its corresponding weighting coefficient, respectively. The pre-trained parameters of SeisDeFM serve as a strong initialization, allowing SeisDeFM to rapidly adapt to specific noise distributions with limited labeled data. Through pre-training and transfer learning, SeisDeFM approximates a universal conditional distribution $P_{\text{denoising}}$ rather than learns a mapping specific to a single noise type:

\begin{equation}
P_{\text{denoising}}\big(x \sim P_{FM}\,|\, y \sim P_{FM}\big).
\label{eq:10}
\end{equation}

Consequently, SeisDeFM exhibits multi-noise adaptability, cross-survey generalization, and efficient transfer learning on unseen datasets. This paradigm shifts seismic denoising from the isolated learning of noise-specific mappings toward universal seismic representations and efficient noise adaptation, forming the theoretical basis for large-scale pre-stack seismic data denoising.

\begin{figure}[htbp]
\centering
\includegraphics[width=\linewidth,height=0.78\textheight,keepaspectratio]{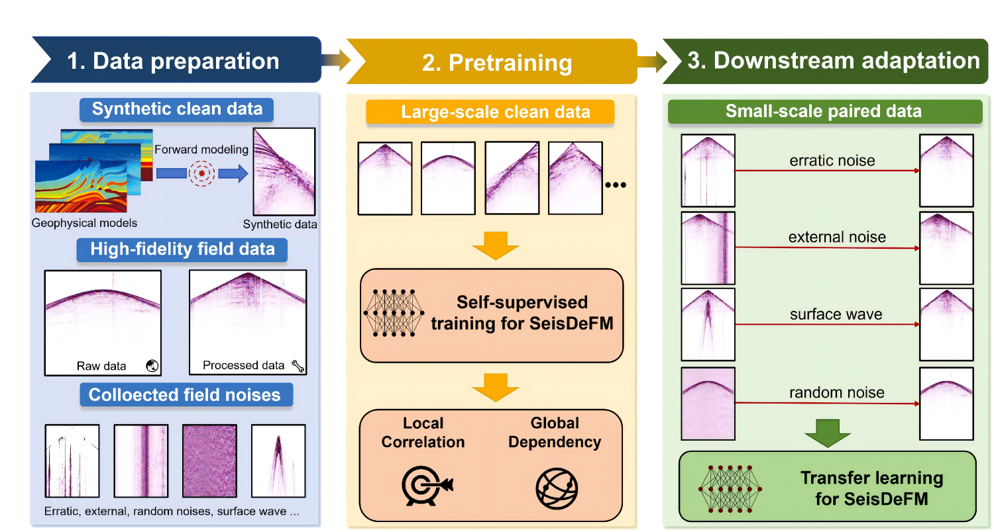}
\caption{The workflow of SeisDeFM.}
\label{fig:1}
\end{figure}

The main workflow of SeisDeFM is shown in Fig.~\ref{fig:1}. In this paper, we present a case study for SeisDeFM that details the preparation of training datasets as well as the strategies for pre-training and downstream adaptation. Furthermore, we demonstrate SeisDeFM's performance by evaluating its denoising effectiveness against four specific types of noise, including erratic noise, external noise, surface waves, and random noise.

\subsection{Data preparation for training SeisDeFM}\label{sec:data-preparation}

Building SeisDeFM with strong generalization capabilities depends fundamentally on constructing high-quality, large-scale, representative datasets. In this section, we detail the data preparation strategies for the self-supervised pre-training and downstream adaptation stages.

The objective of the pre-training stage is to enable SeisDeFM to comprehensively learn the universal geophysical laws of seismic wavefields and the spatial correlations of seismic events. Therefore, sufficient clean seismic data are required for pre-training. Synthetic clean data can be obtained via wave-equation-based forward modeling, whereas clean field data are derived from the nearly clean raw data or the data processed with conventional processing approaches. The detailed methods for constructing clean seismic datasets are as follows:

(1) Forward modeling based on wave equations. Forward modeling involves establishing geophysical models for the subsurface medium (velocity, density, Q factor and Thomsen parameters etc.), setting an appropriate observation system (towed marine streamer, ocean-bottom cable (OBC) and vertical seismic profile (VSP) etc.), and performing numerical simulation with corresponding wave equations (acoustic, elastic, viscoelastic and anisotropic etc.) to produce clean seismic records. Taking the isotropic acoustic wave equation as an example, we implement the forward modeling process as follows:

\begin{equation}
\frac{1}{v^{2}(\boldsymbol{r})}
\frac{\partial^{2}u(\boldsymbol{r},t)}{\partial t^{2}}
=
\nabla^{2}u(\boldsymbol{r},t)+f(\boldsymbol{r},t).
\label{eq:11}
\end{equation}

where u, v and f represent the wavefield, velocity model and source function, respectively, r denotes the coordinate vector in 3-dimensional (3D) space. The primary advantage of this method lies in its ability to acquire absolutely clean data with known ground-truth labels. However, numerical simulations are often overly idealized and struggle to perfectly replicate the complex medium characteristics and true wavefield responses inherent in field data. Consequently, models pre-trained solely on synthetic data often suffer from generalization limitations when applied to field data.

(2) Field data with high-fidelity. This approach directly utilizes massive volumes of field seismic data for training. For seismic acquisition carried out under favorable environmental conditions with simple subsurface structures and no complex near-surface interferences, the raw shot gathers or data in partial spatio-temporal regions exhibit relatively high SNR and can be adopted as clean data. However, the raw clean data requires stringent conditions, and such data are therefore scarce in quantity. For raw data containing a small amount of noise, the data denoised by conventional methods can be used as the clean field data for pre-training. The clean field data preserve to the greatest extent the complex kinematic and dynamic characteristics of field seismic wavefields, ensuring that the features extracted by SeisDeFM adhere strictly to geophysical laws. However, refining raw field data into high-fidelity clean data demands substantial manual effort from domain experts as well as rigorous validations. This procedure renders such high-quality field datasets exceedingly scarce and costly to acquire.

During the downstream adaptation stage for specific denoising tasks, it is necessary to construct paired noisy-clean datasets for supervised transfer learning. To maximize SeisDeFM's adaptability to complex field conditions, target noises must be added into the clean data. The mainstream practice involves extracting real noise directly from field seismic data. To enhance the robustness of the downstream tasks against highly variable noise distributions, an augmentation strategy for different kinds of noises by generative models can be adopted \citep{dongx2022}.

\begin{figure}[htbp]
\centering
\includegraphics[width=\linewidth,height=0.78\textheight,keepaspectratio]{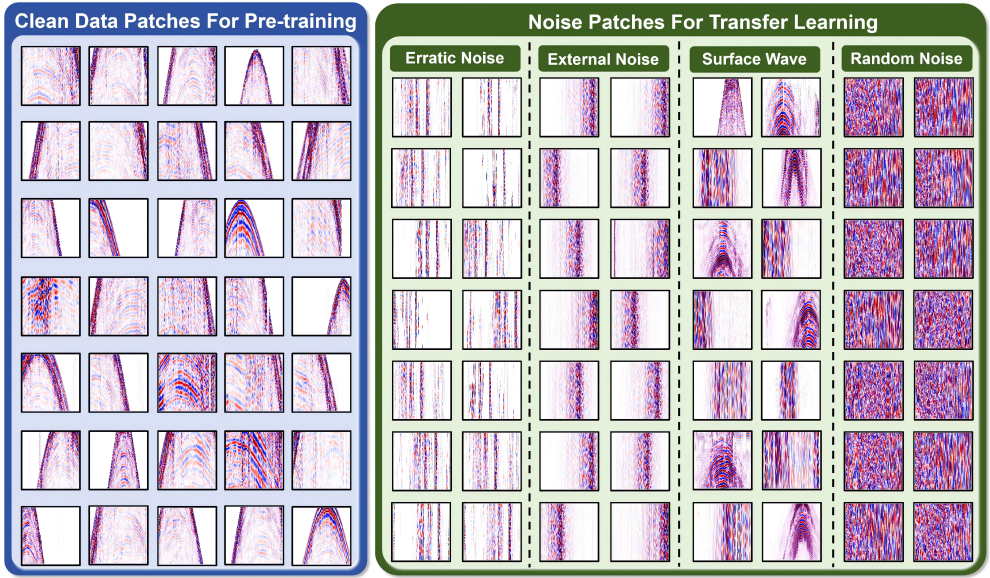}
\caption{Features of clean data patches for SeisDeFM pre-training and features of noise patches for SeisDeFM transfer learning.}
\label{fig:2}
\end{figure}

In this paper, to achieve superior denoising performance, we adopt the clean field data as the pre-training dataset for SeisDeFM to mitigate domain-shift issues arising from synthetic clean data. Meanwhile, real noise is added to clean field data to build paired datasets for transfer learning in downstream tasks. As illustrated in Fig.~\ref{fig:2}, the clean data contain abundant local event features, and the real noise covers a highly diverse set of complex field interference patterns, guaranteeing sufficient diversity in both clean-signal and noise distributions. During training, using a complete shot gather as a single input channel greatly helps SeisDeFM capture global wavefield dependencies, whereas cropping a shot gather into smaller patches mainly serves to reduce computational costs. To obtain better denoising performance, the case study presented in this paper uses the strategy of taking the complete shot gather as a single-channel input.

\subsection{The architecture of SeisDeFM}\label{sec:architecture}

\begin{figure}[htbp]
\centering
\includegraphics[width=\linewidth,height=0.78\textheight,keepaspectratio]{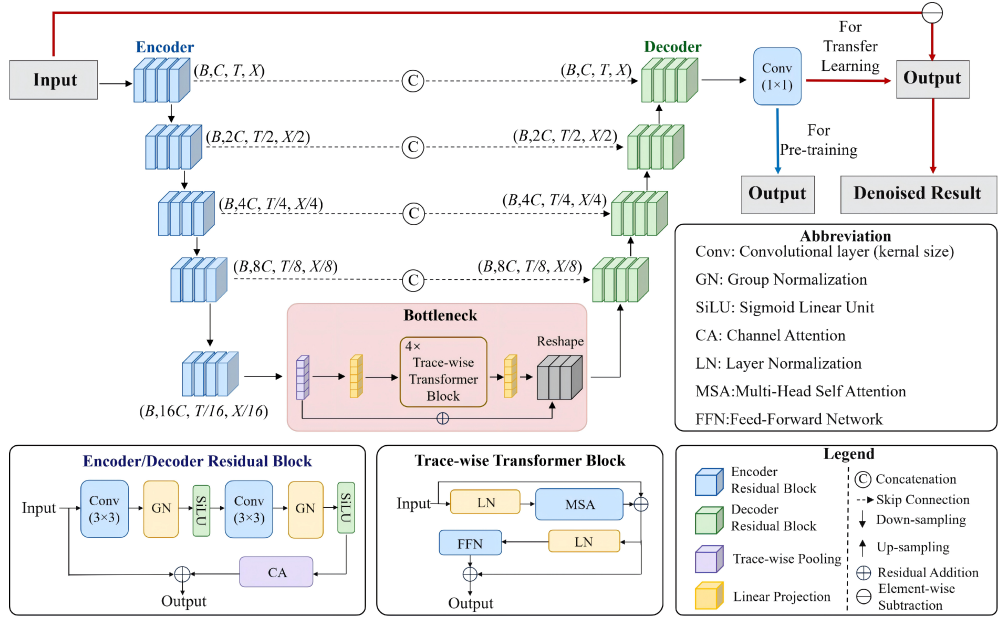}
\caption{The detailed architecture of SeisDeFM.}
\label{fig:3}
\end{figure}

In recent years, foundation-model architectures including Transformer, Mamba, and diffusion models have exhibited revolutionary performance in diverse fields, benefiting from their powerful global context modeling and capacity for generating complex data distributions \citep{kumara2024,chengs2025,dongx2026}. Nevertheless, naive deployment of these generic architectures on seismic data cannot fully characterize the unique spatio-temporal relationships and strong inter-trace coherence embedded in seismic wavefields. To tackle this limitation, we present a SeisDeFM architecture tailored for seismic denoising in our case study (Fig.~\ref{fig:3}).

The network adopts a residual U-Net backbone to extract local multi-scale features, alongside a bottleneck equipped with a trace-wise Transformer for global information fusion. Such a design trades off the U-Net's strength in retaining local seismic events and the Transformer's superiority in capturing long-range trace dependencies, thereby optimizing SeisDeFM for space-time-domain signal-noise separation. The internal data flow of SeisDeFM consists of four key stages:

(1) Local multi-scale feature encoding. The input seismic data for SeisDeFM can be denoted as $ Y\in\mathbb{R}^{B\times C\times T\times X}$, where B, C, T and X represent the batch size, number of channels, number of time samples, and number of traces, respectively. The U-Net encoder consists of five successive residual convolutional blocks and max-pooling layers. Each residual block includes 3 $\times$ 3 convolutions, group normalization, Sigmoid-Linear Unit (SiLU) activation functions, and a channel attention mechanism. Skip connections are adopted within each block to mitigate the vanishing-gradient problem in deep networks. By progressively reducing spatial resolution and enlarging the receptive field, the encoder extracts multi-scale local features of seismic waveforms.

(2) Trace-wise global feature extraction at the bottleneck. To mitigate the limited receptive field inherent to the convolution-based U-Net, SeisDeFM integrates a bottleneck module equipped with Transformer blocks to extract global features. The bottleneck's input feature maps undergo temporal pooling and are split into sequence tokens along the seismic-trace dimension. Inside the trace-wise Transformer blocks, each token generates query (Q), key (K), and value (V) through linear projections. Multi-head self-attention (MSA) is subsequently applied to model dependencies among distinct seismic traces:

\begin{equation}
\text{MSA}(Q,K,V)\equiv \mathrm{softmax}\left(\frac{QK^\mathrm{TP}}{\sqrt{d_k}}\right)V,
\label{eq:12}
\end{equation}

where $d_k $ denotes the dimension of the keys, and $\mathrm{TP}$ represents the transposition operation. This mechanism allows the network to adaptively exploit global context from distant seismic traces for learning both noise patterns and signal features. The resulting global information is then fused with local CNN-derived features through residual connections, yielding robust seismic representations.

(3) The architecture of the decoder. The decoder gradually recovers the spatio-temporal resolution of bottleneck feature maps using transposed convolutions. In every upsampling step, skip connections concatenate encoder-derived shallow features of the matching scale with the current high-dimensional features along the channel dimension. Such a design offsets the irreversible spatial-detail loss caused by downsampling and enables effective reconstruction of input data within the space-time domain.

(4) The output data of SeisDeFM. The feature maps produced by the decoder are mapped to a single-channel feature map via a 1 $\times$ 1 convolutional layer. During the self-supervised pre-training stage, the training objective of SeisDeFM typically involves full-wavefield reconstruction of masked incomplete input data and the maximization of the distance between positive and negative sample pairs for contrastive learning. Accordingly, the output of the 1 $\times$ 1 convolutional layer can directly serve as the final prediction in the pre-training stage, as indicated by the blue arrow. In the transfer-learning stage, requiring the pre-trained model to directly predict clean seismic signals for downstream denoising tasks frequently leads to convergence difficulties owing to the complex characteristics of seismic wavefields. To tackle this issue, the training objective is switched to predicting the noise component to be removed. A global residual connection is built between the original noisy input and the predicted noise, as indicated by the red arrows. The final denoised seismic data are obtained by subtracting the predicted noise from the original noisy input:

\begin{equation}
\hat{x}=y-\hat{n},
\label{eq:13}
\end{equation}

where $\hat{n}$ and $\hat{x}$ represent the predicted noise and denoised result, respectively.

\subsection{The pre-training of SeisDeFM}\label{sec:pretraining}

\begin{figure}[htbp]
\centering
\includegraphics[width=\linewidth,height=0.78\textheight,keepaspectratio]{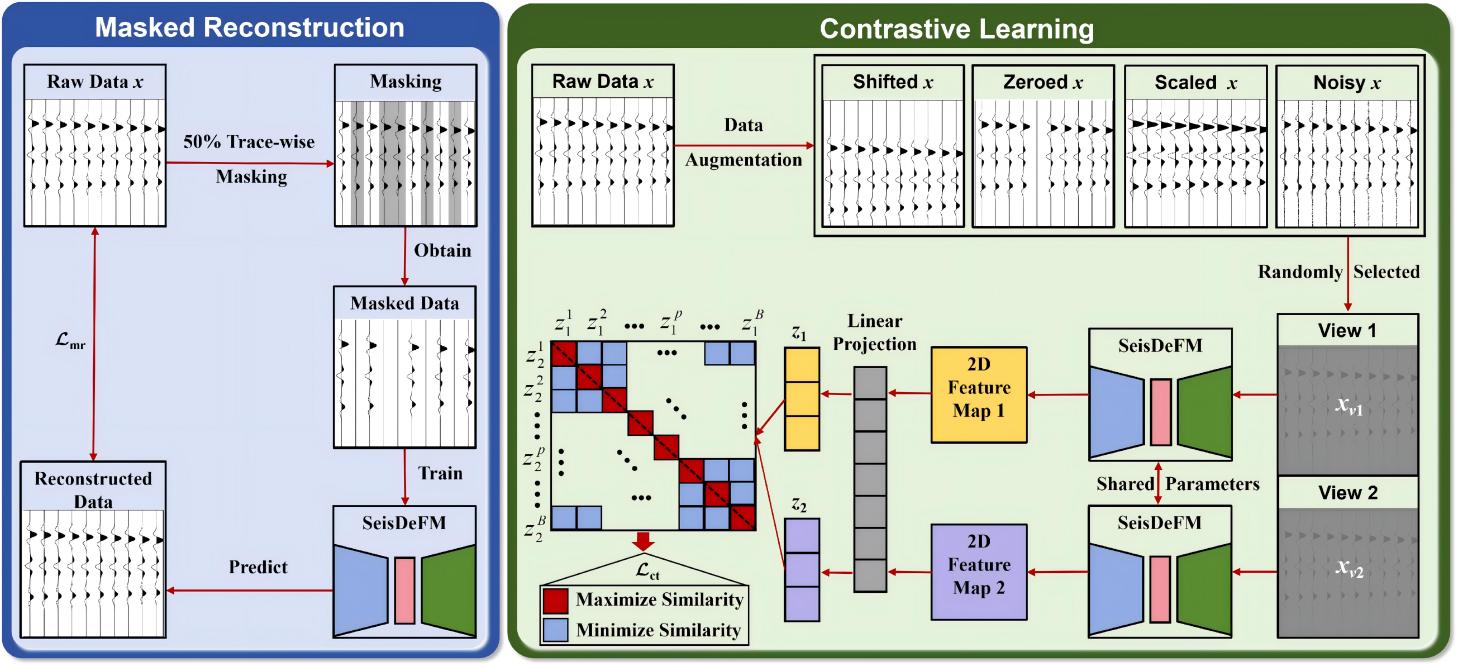}
\caption{The joint pre-training strategy of SeisDeFM.}
\label{fig:4}
\end{figure}

At present, self-supervised pre-training schemes for foundation models mainly fall into two categories: generative approaches (including generative models, reconstructive methods, and autoregressive methods) and contrastive approaches (including context-instance and instance-instance contrasts) \citep{liuq2025}. Inspired by these existing paradigms, our case study implements a joint self-supervised learning strategy for SeisDeFM that combines masked wavefield reconstruction with contrastive learning (Fig.~\ref{fig:4}). This scheme targets the simultaneous capture of local seismic-event continuity and global wavefield correlations. In this pre-training stage, the masked reconstruction and contrastive learning branches receive identical copies from the input batch while sharing all network parameters.

To enable SeisDeFM to learn the principles of seismic wavefields, especially the inter-trace dependencies, we employ a trace-wise masking strategy that better conforms to field seismic scenarios with missing traces or sparse sampling. Specifically, 50\% of the traces in the input seismic data are randomly masked to generate the incomplete data. The model is tasked with reconstructing the complete data from this incomplete input. The reconstruction loss function for clean seismic data is defined as the mean squared error (MSE) computed only over the masked traces:

\begin{equation}
\mathcal{L}_{mr}\equiv \frac{1}{L}\sum_{l\in L}\left\|\mathcal{F}_\Theta(x_{\text{mask}})_l - x_l\right\|_2^2,
\label{eq:14}
\end{equation}

where $\mathcal{L}_{mr}$ represents the mask-reconstruction loss function,  $ x_{\text{mask}}$ denotes the masked seismic data, and $ a $ denotes the index set of the masked traces with the maximal number of $ A $.

To further enhance the ability of SeisDeFM to represent the global characteristics of clean seismic data, we introduce a contrastive learning mechanism. Given the input clean seismic data $ x $, we generate two distinct augmented views $ x_{v1}$ and $ x_{v2}$ by randomly selecting and combining two out of the following four data augmentation strategies:

(1) All seismic traces are shifted by 5 samples along the time axis, either in the forward or backward direction.

(2) Randomly setting 5\% of the seismic traces to zero which is similar to the operation of trace-wise masking.

(3) Scaling amplitudes by multiplying the seismic traces with a random coefficient uniformly sampled between 0.8~1.2.

(4) Adding slight Gaussian noise with a standard deviation:

\begin{equation}
\sigma_G \equiv 0.02 \times \sigma_x,
\label{eq:15}
\end{equation}

where $\sigma_G $ and $\sigma_x $ represent the standard deviation of the Gaussian noise and the seismic data to be added noise. This process adaptively adds noise based on the statistical characteristics of the corresponding input seismic data, preventing excessive noise from covering effective signals.

The two augmented views are processed by SeisDeFM with shared network parameters. To compute the contrastive loss, a linear projection head is appended to the output of the 1$\times $ 1 convolutional layer to flatten and project the 2-dimensional (2D) feature maps into 1-dimensional (1D) latent vectors:

\begin{equation}
z_1 = H\big[\mathcal{F}_\Theta(x_{v1})\big] \text{ and } z_2 = H\big[\mathcal{F}_\Theta(x_{v2})\big],
\label{eq:16}
\end{equation}

where $z_1 $ and $ z_2 $ represent the 1D latent vectors after linearly projecting, and $ G(\bullet)$ represents the operator of linear projection. Specifically, the linear projection head is implemented as a two-layer multi-layer perceptron (MLP), with a rectified linear unit (ReLU) activation sandwiched between the two-layer of MLP.

Contrastive learning employs an information noise contrastive estimation (InfoNCE) loss function. InfoNCE first calculates the pairwise feature similarity matrix between the two augmented views. Then, it treats the two augmented views derived from the same sample as a positive pair, and views from all other samples within the batch as negative samples. Finally, it optimizes the relative similarity between positive and negative samples via a cross-entropy loss formulation:

\begin{equation}
\mathcal{L}_{ct} = -\frac{1}{B}\sum_{p=1}^{B}\log\frac{\exp\big[\mathit{sim}(z_1^p,z_2^p)/\tau\big]}{\sum_{q=1}^{B}\exp\big[\mathit{sim}(z_1^p,z_2^q)/\tau\big]},
\label{eq:17}
\end{equation}

where $\mathcal{L}_{ct}$ represents the InfoNCE loss function, p and q denote the indices of the query sample and the comparison sample within the batch, respectively, $ z_1^p $and $ z_2^p $represent the 1D latent vectors of the two augmented views generated from the $ p^{\mathrm{th}}$ seismic data in a batch which composes a positive pair,$ z_2^q $represents the 1D latent vector of the augmented view from the $ q^{\mathrm{th}}$ sample which is used to compose a negative pair with when p $\neq $ q, $\tau $ is a temperature parameter used to scale the similarity scores and regulate the model's penalty for hard negative samples which is set to 0.1 in this study, and $\mathit{sim}(\cdot)$ denotes the cosine similarity function.

The overall pre-training loss function of SeisDeFM is the weighted sum of the masked reconstruction loss and the contrastive learning loss:

\begin{equation}
\mathcal{L}_{pt} \equiv \mathcal{L}_{mr} + \gamma \mathcal{L}_{ct},
\label{eq:18}
\end{equation}

where $\mathcal{L}_{pt}$ represents the complete pre-training loss function, and $\gamma $ is a weighting factor that adjusts the relative importance of the two loss terms. Within this joint optimization framework, the masked reconstruction primarily focuses on capturing the local continuity and inter-trace features of seismic events, and the contrastive learning emphasizes learning the global characteristics and long-range contextual representations of the entire seismic data. Consequently, this pre-training strategy efficiently equips SeisDeFM with a robust, generalizable feature representation that serves as a powerful initialization for downstream denoising tasks.

\subsection{Transfer learning of SeisDeFM for downstream tasks}\label{sec:transfer-learning}

Transfer learning of foundation models for downstream tasks fundamentally aims to adapt pre-trained parameters to specific task distributions at minimal computational cost while preserving the intrinsic learned capabilities. Transfer-learning approaches can be broadly categorized into full-parameter fine-tuning, parameter-efficient fine-tuning (PEFT) techniques (including additive and selective fine-tuning, reparameterized fine-tuning, and hybrid fine-tuning), instruction tuning, multi-task joint fine-tuning, unsupervised domain adaptation, and knowledge-distillation-based transfer learning. Considering the strong non-stationarity of seismic wavefields across heterogeneous subsurface media and the complexity of field noise, we present a case of downstream-task adaptation for SeisDeFM, in which a full-parameter fine-tuning strategy is adopted.

\begin{figure}[htbp]
\centering
\includegraphics[width=\linewidth,height=0.78\textheight,keepaspectratio]{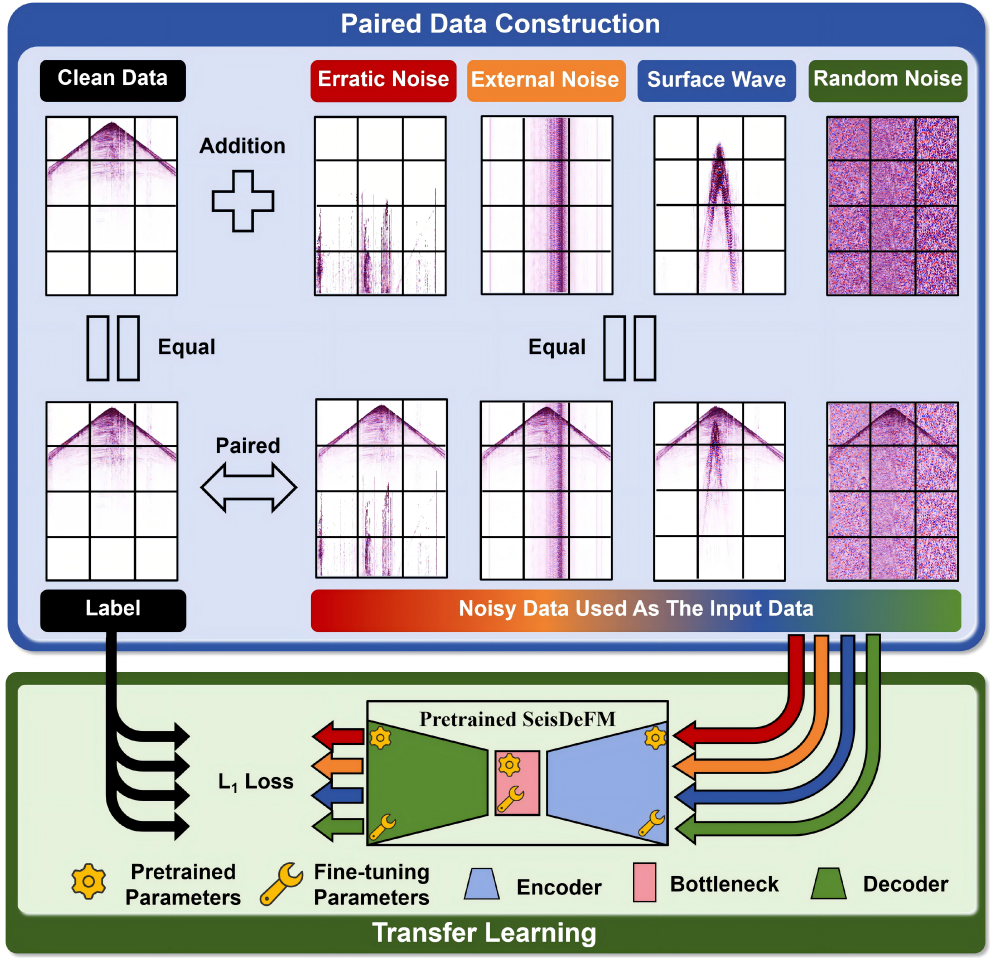}
\caption{The downstream adaptation strategy of SeisDeFM.}
\label{fig:5}
\end{figure}

During the transfer-learning stage, the network parameters obtained from pre-training serve as a robust initialization for SeisDeFM. To adapt the model to the specific denoising task, we update all model parameters using a small-scale paired dataset, as formulated in Equation~\eqref{eq:8}. The denoised results are obtained via element-wise wavefield subtraction, as illustrated in Equation~\eqref{eq:13}. The optimal parameters of SeisDeFM are solved for under the L1-norm constraint for downstream denoising tasks:

\begin{equation}
\widehat{\Theta}=\arg\min_{\Theta}\frac{1}{M}\sum_{i=1}^{M}\big\|F_{\Theta}\big[y_i(h)\big]-x_i\big\|_1,
\label{eq:19}
\end{equation}

By leveraging the inherent wavefield characteristics captured during pre-training, this full-parameter fine-tuning strategy enables rapid alignment with the target noise distribution while effectively mitigating the risk of overfitting on small-scale downstream datasets. The downstream adaptation strategy adopted in our SeisDeFM case study is illustrated in Fig.~\ref{fig:5}. To comprehensively demonstrate the efficient downstream adaptation capabilities, we select erratic noise, external noise, surface waves and random noise as downstream evaluation cases. These interference components represent the most typical seismic interferences, ranging from ubiquitous incoherent energies and localized high-amplitude anomalies to strongly coherent dispersive wavefields, thereby rigorously validating the universal multi-noise adaptability of SeisDeFM.

\subsection{The evaluation metrics for SeisDeFM}\label{sec:evaluation-metrics}

To comprehensively evaluate the performance of SeisDeFM, multiple evaluation metrics should be employed, which can be broadly categorized into computational efficiency metrics and quantitative denoising quality metrics.

Evaluating SeisDeFM requires a detailed accounting of both data requirements and computational cost. The following metrics are usually used to quantify the efficiency of SeisDeFM:

(1) Data volume and training time. We record the volume of training data, including the clean data used for pre-training and the paired clean--noisy data used for transfer learning, as well as the corresponding training time for both the self-supervised pre-training stage and the downstream adaptation stage. A robust SeisDeFM should demonstrate that pre-training with massive data significantly shortens convergence time and reduces dependency on data volume during transfer learning.

(2) Trainable parameters within SeisDeFM. This metric counts the total number of trainable weights and biases in the SeisDeFM architecture. It reflects the memory footprint and the feasibility of deploying SeisDeFM on computationally constrained platforms.

(3) Floating-point operations. Floating-point operations (FLOPs) measure the total number of mathematical operations required to process input seismic data. This metric serves as a direct indicator of SeisDeFM's computational complexity and inference speed, which is vital for large-scale industrial seismic data processing.

To evaluate the fidelity of signal--noise separation achieved by SeisDeFM denoising, the following four core quantitative metrics can be adopted for this purpose:

(1) Mean absolute error (MAE) calculates the average absolute difference between the denoised and the clean seismic data:

\begin{equation}
\mathrm{MAE}=\frac{1}{M}\sum_{i=1}^{M}\big|\widehat{x}_i-x_i\big|,
\label{eq:20}
\end{equation}

(2) Root-mean-square error (RMSE) measures the standard deviation of residual errors:

\begin{equation}
\mathrm{RMSE}=\sqrt{\frac{1}{M}\sum_{i=1}^{M}\big(\widehat{x}_i-x_i\big)^2}.
\label{eq:21}
\end{equation}

By squaring the amplitude differences, RMSE heavily penalizes larger local errors, effectively reflecting the residual noise energy remaining in the denoising result.

(3) Peak signal-to-noise ratio (PSNR) evaluates the logarithmic ratio of the maximum possible signal power to the mean squared error of the signal distortion:

\begin{equation}
\mathrm{PSNR}=10\log_{10}\left[\frac{\max(x)^2}{\mathrm{MSE}}\right],
\label{eq:22}
\end{equation}

where $\mathrm{MSE}=\mathrm{RMSE}^2 $, and $\max(\cdot)$ denotes the operator that calculates the peak amplitude of the clean seismic data. A higher PSNR value indicates a higher degree of overall signal fidelity and superior noise suppression.

(4) Structural similarity index measure (SSIM) measures the structural coherence between the denoised and clean data by comparing local patterns of pixel intensities:

\begin{equation}
\mathrm{SSIM}(x,\widehat{x})=\frac{\big(2\mu_x\mu_{\widehat{x}}+c_1\big)\big(2\sigma_{x\widehat{x}}+c_2\big)}{\big(\mu_x^2+\mu_{\widehat{x}}^2+c_1\big)\big(\sigma_x^2+\sigma_{\widehat{x}}^2+c_2\big)},
\label{eq:23}
\end{equation}

where $\mu_x$ and $\mu_{\widehat{x}}$ are the mean values of clean data and denoised result, respectively, $\sigma_x^2 $ and $\sigma_{\hat{x}}^2 $ are the variances of clean data and denoised result, respectively, $\sigma_{x\widehat{x}}$ denotes the covariance between the clean data and the denoised result, and $ c_1 $, $ c_2 $ are small constants to stabilize the division. A higher SSIM directly corresponds to the better preservation of local features of seismic events.

\section{A case study of SeisDeFM}\label{sec:case-study}

To validate the effectiveness of SeisDeFM, a series of experiments were conducted under different noise conditions. The noisy seismic data used in this study were collected from a field seismic survey in China. The complete dataset contains 12,000 samples, including 10,000 clean-noisy paired training samples, 1,000 validation samples, and 1,000 testing samples. All data used in this case study have a sampling interval of 0.002 s and a trace spacing of 25 m. In our setup, SeisDeFM was first pre-trained on all clean data from the full training dataset for 30 epochs. The pre-trained weights were then adopted to initialize SeisDeFM for the downstream denoising task. During the subsequent supervised transfer training stage, only 1,000 paired samples were randomly selected from the training set for another 30-epoch training. U-Net and Transformer baselines, which belong to TSDM, were adopted as competing models and trained on the complete training dataset for 30 epochs. The adopted Transformer baseline uses four attention heads in the multi-head attention module, and four Transformer blocks. The adopted U-Net baseline consists of a four-layer encoder and a four-layer decoder, where each layer contains two 3$\times $ 3 convolution layers and two 1$\times $ 1 channel attention convolution layers. All models in this study were trained on the same NVIDIA A10 GPU with a batch size of 8. The AdamW optimizer was employed for training with an initial learning rate of $1\times 10^{-4}$, and the learning rate remained fixed throughout training, with no additional decay schedule applied.

\subsection{The analysis of training efficiency}\label{sec:training-efficiency}

Since the dataset size and experimental environment are identical for all experiments in this case, the computational complexity of each model does not vary across different downstream denoising tasks. Accordingly, this section compares and analyzes the computational complexity and training efficiency of SeisDeFM and TSDM. Table~\ref{tab:1} shows that the U-Net baseline contains 10.67~M trainable parameters and requires 341.14~GFLOPs. Its convolution-dominated encoder-decoder architecture enables repeated multiscale feature extraction and dense spatial aggregation, resulting in the longest supervised training time of 1,538~min. The Transformer baseline adopted in this study contains 14.52~M trainable parameters and requires 38.40 GFLOPs. Although the Transformer and U-Net baselines have comparable numbers of trainable parameters, Transformer requires substantially fewer FLOPs due to its more efficient feature representation, while its training time remains comparable because self-attention involves computationally intensive query--key similarity calculations and attention-weighted feature aggregation, which introduce substantial memory and computational overhead. SeisDeFM has 12.84~M trainable parameters and consumes 604~GFLOPs in the pre-training stage. After pre-training, SeisDeFM in the transfer-learning stage has 12.51~M trainable parameters and exhibits a computational cost comparable to the U-Net baseline with 341.20~GFLOPs. These results demonstrate that the additional Transformer-based bottleneck introduce only a modest increase in model size and computational overhead. More importantly, SeisDeFM requires merely 1,000 paired samples for transfer learning, shortening its supervised training time to 108~min. Therefore, although SeisDeFM is not the lightest architecture, its pre-training computational cost is comparable to those of TSDM models. It further supports fast and lightweight transfer learning for noise-specific denoising tasks using a small number of paired samples, achieving accurate denoising under multiple noise types.


\begin{table}[htbp]
\centering
\caption{Quantitative comparison of training efficiency.}
\label{tab:1}

\small
\setlength{\tabcolsep}{4.0pt}
\renewcommand{\arraystretch}{1.22}

\begin{tabularx}{\linewidth}{
@{}
>{\raggedright\arraybackslash}X
>{\centering\arraybackslash}m{0.14\linewidth}
>{\centering\arraybackslash}m{0.14\linewidth}
>{\centering\arraybackslash}m{0.15\linewidth}
>{\centering\arraybackslash}m{0.11\linewidth}
@{}
}

\toprule

\multicolumn{1}{c}{
    \makecell[c]{\textbf{Method}}
}
&
\makecell[c]{
    \textbf{Training}\\
    \textbf{Data}\\
    \textbf{Volume}
}
&
\makecell[c]{
    \textbf{Training}\\
    \textbf{Time}\\
    \textbf{(min)}
}
&
\makecell[c]{
    \textbf{Trainable}\\
    \textbf{Parameters}\\
    \textbf{(M)}
}
&
\makecell[c]{
    \textbf{GFLOPs}
}
\\

\midrule

U-Net
& 10,000
& 1,538
& 10.67
& 341.14
\\

Transformer
& 10,000
& 1,204
& 14.52
& 38.40
\\

\shortstack[l]{SeisDeFM\\(pre-training)}
& 10,000
& 1,396
& 12.84
& 604.00
\\

\shortstack[l]{SeisDeFM\\(transfer learning)}
& 1,000
& 108
& 12.51
& 341.20
\\

\bottomrule

\end{tabularx}
\end{table}

\subsection{Erratic Noise Attenuation}\label{sec:erratic-noise-results}

Erratic noise corresponds to sparse, non-Gaussian, impulsive random anomalies during field seismic data acquisition. It generally manifests as locally isolated high-amplitude perturbations that can mask weak reflections. To evaluate the denoising capability of SeisDeFM for noisy seismic data contaminated with erratic noise, we compared the denoised results obtained by Transformer,  U-Net and SeisDeFM under identical training conditions.

\begin{figure}[htbp]
\centering
\includegraphics[width=\linewidth,height=0.6\textheight,keepaspectratio]{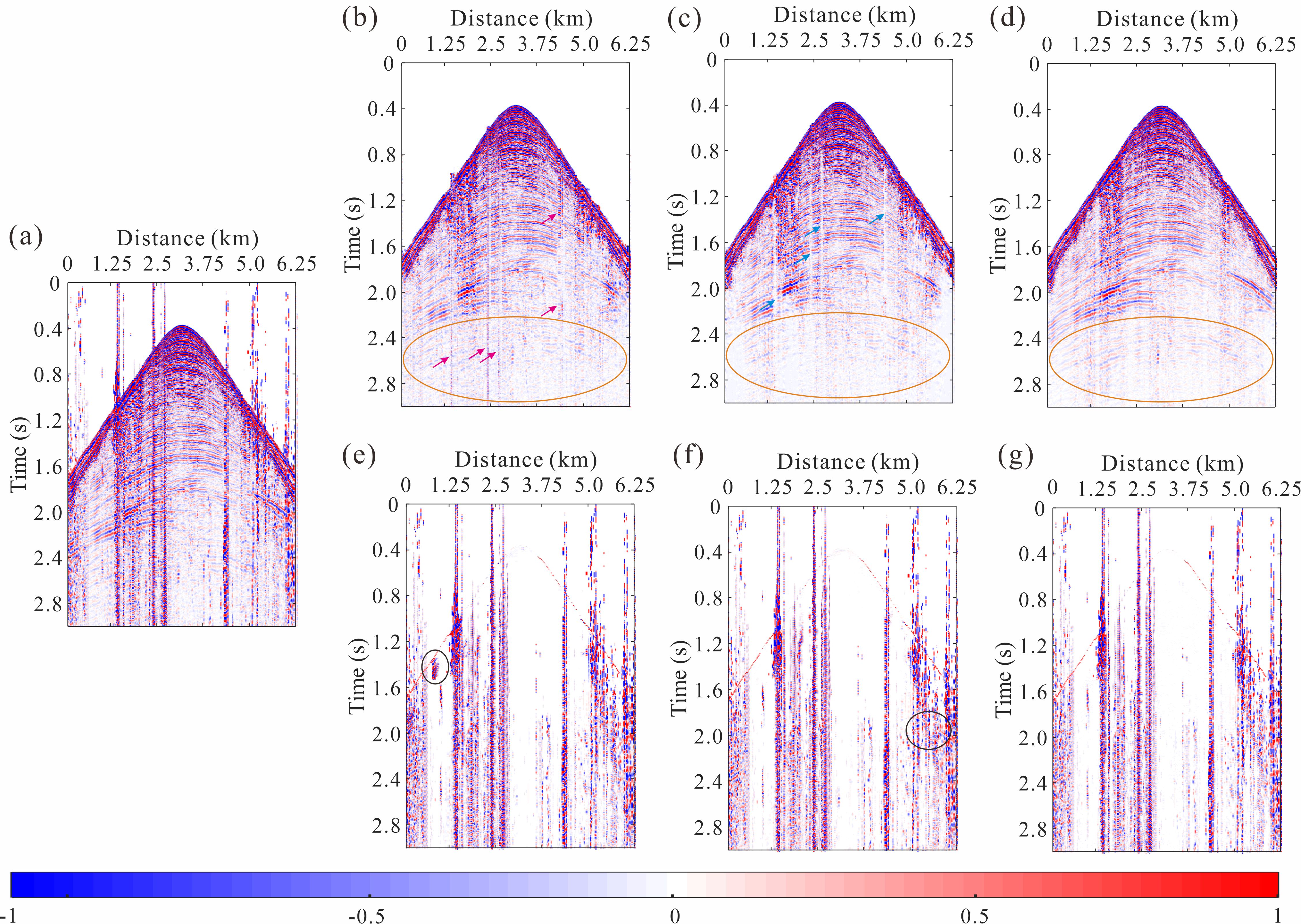}
\caption{Comparisons of erratic-noise attenuation. (a) Raw noisy input data, (b)-(d) denoised results of Transformer, U-Net, and SeisDeFM, (e)-(g) the residual profiles between the input data shown in (a) and the denoised results shown in (b)-(d).}
\label{fig:6}
\end{figure}

A representative sample of raw noisy data was selected for comparison, as shown in Fig. \ref{fig:6}(a). Fig. \ref{fig:6}(b) shows the denoised result obtained by Transformer, and Fig. \ref{fig:6}(e) shows its residual profile. As highlighted by the red arrows in Fig. \ref{fig:6}(b), Transformer fails to completely eliminate all the high-amplitude erratic noise, with residual noise remaining at different time instants. This phenomenon is mainly caused by the lack of local waveform constraints in the global representation of self-attention, where effective reflections and sparse noise perturbations may be mixed in the global attention space. The corresponding residual map shown in Fig. \ref{fig:6}(e) further reveals that, in addition to incomplete noise suppression, Transformer suffers from locally continuous signal leakage as indicated by the black circle. Fig. \ref{fig:6}(c) shows the denoised result of U-Net. Compared with Transformer, U-Net achieves better removal of the high-amplitude erratic noise. Its convolution-based local feature extraction performs well in suppressing the erratic noise characterized by abnormal amplitude variations. However, as indicated by the blue arrows in Fig. \ref{fig:6}(c), some of the seismic signals are also removed during denoising. In addition, the black circle in Fig. \ref{fig:6}(f) indicates that the signal leakage exists in the residual profile. This is because the limited receptive field of convolutional operations prevents U-Net from fully capturing long-range dependencies and the global continuity of seismic events, causing weak reflections to be misidentified as noise. In contrast, SeisDeFM produces the cleanest denoised result (Figs. \ref{fig:6}(d) and \ref{fig:6}(g)). In particular, as highlighted by the yellow circles in Figs. \ref{fig:6}(b)-\ref{fig:6}(d), both the Transformer and U-Net baselines exhibit obvious residual noise and signal loss in the denoised results of late arrivals. In contrast, SeisDeFM preserves the continuity and amplitude of events for deep weak signals while removing noise effectively. By combining the local feature extraction capability of U-Net with the long-range dependency modeling of Transformer, and by introducing masked reconstruction and contrastive learning during pre-training, SeisDeFM better distinguishes erratic noises from physically meaningful seismic events. As a result, it suppresses strong noise while maintaining event continuity. Notably, SeisDeFM uses only 1,000 transfer-learning samples and requires the shortest training time (Table~\ref{tab:1}), while yielding the best denoising performance.

\begin{figure}[htbp]
\centering
\includegraphics[width=\linewidth,height=0.6\textheight,keepaspectratio]{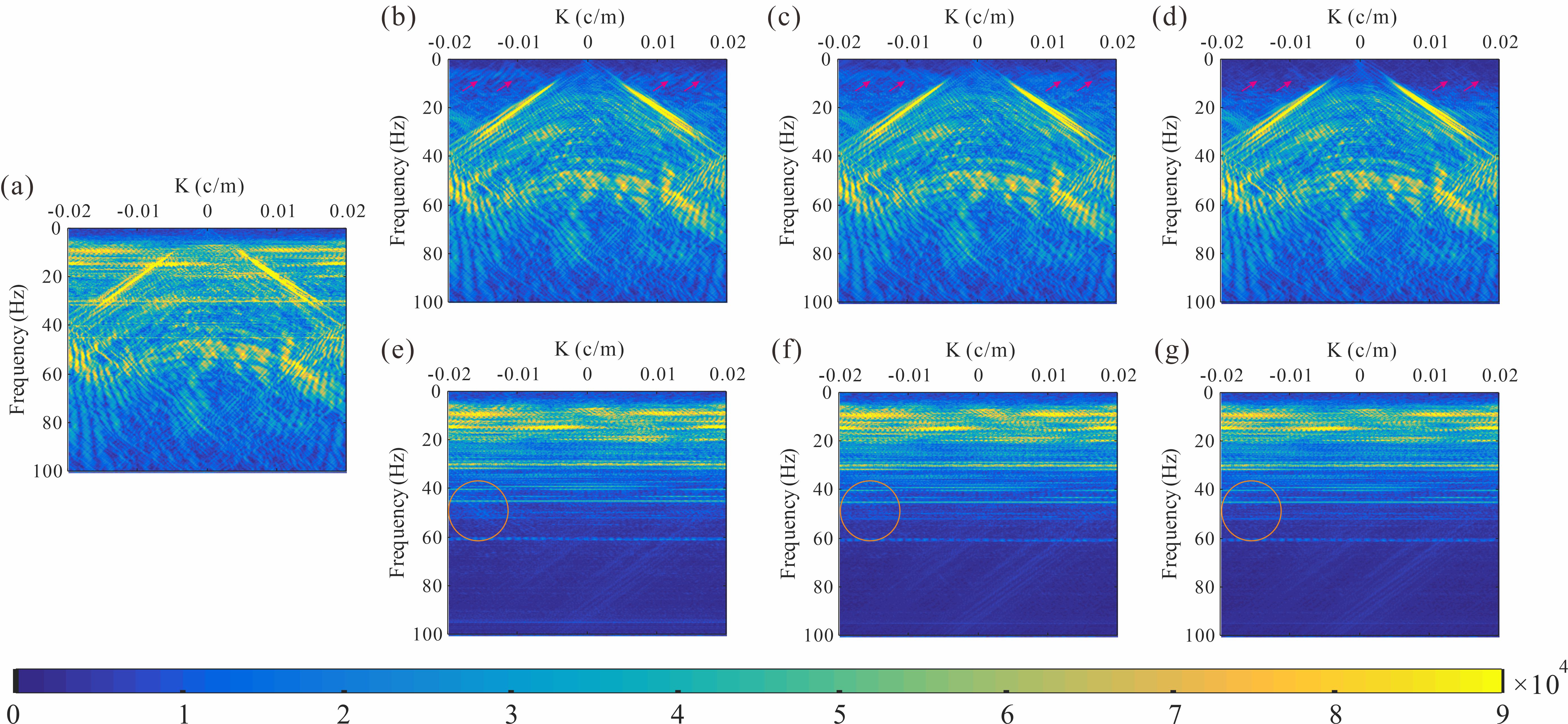}
\caption{F-K spectra comparisons of erratic noise attenuation. (a)-(g) F-K spectra of the data shown in Figs. \ref{fig:6}(a)-\ref{fig:6}(g).}
\label{fig:7}
\end{figure}

We further conducted F-K spectral analysis on the data shown in Fig.~\ref{fig:6}, as presented in Fig.~\ref{fig:7}. As indicated by the red arrows in Figs. \ref{fig:7}(b)-\ref{fig:7}(d), obvious low-frequency residual noise exists in the denoised results of Transformer and U-Net, whereas SeisDeFM achieves favorable separation of low-frequency signals and noise. Meanwhile, as indicated by the yellow circles in Figs. \ref{fig:7}(e) and \ref{fig:7}(f), signal leakage can be observed in the denoised results of Transformer and U-Net. However, this limitation is effectively alleviated by SeisDeFM as illustrated by Fig \ref{fig:7}(g). These F-K spectra demonstrate that SeisDeFM further improves erratic-noise suppression and maintains the spectral integrity of seismic signals.

\subsection{External Noise Attenuation}\label{sec:external-noise-results}

External noise refers to interference introduced by environmental or human activities during seismic acquisition, such as vehicle movement, mechanical vibration, electrical interference, and other nearby operations. This type of noise may exhibit coherent characteristics and can substantially degrade the quality of field seismic records.

\begin{figure}[htbp]
\centering
\includegraphics[width=\linewidth,height=0.6\textheight,keepaspectratio]{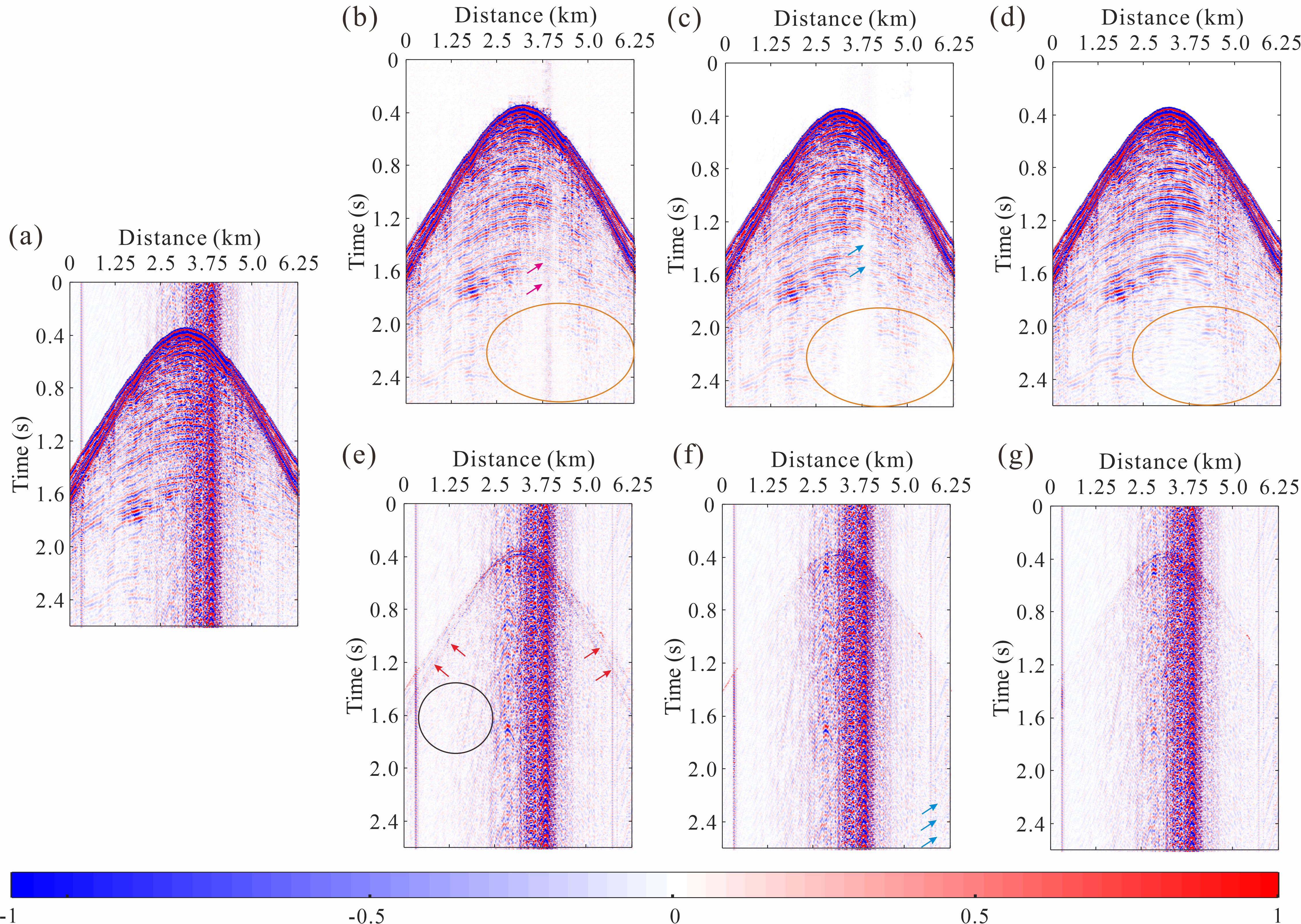}
\caption{Comparisons of external-noise attenuation. (a) Raw noisy input data, (b)-(d) denoised results of Transformer, U-Net, and SeisDeFM, (e)-(g) the residual profiles between the input data shown in (a) and the denoised results shown in (b)-(d).}
\label{fig:8}
\end{figure}

The denoised results for a raw noisy data are shown in Fig.~\ref{fig:8}. As can be seen in Figs. \ref{fig:8}(b),  strong residual external-noise can be observed in the denoised results of Transformer, as indicated by the red arrows. Meanwhile, the residual profile of Transformer shows strong leakage of early arrivals (as indicated by the red arrows) and continuous reflections at later time (as indicated by the black circle). This reveals that the global representation of the Transformer baseline is insufficient to distinguish external noise from weak but continuous seismic reflections. Fig. \ref{fig:8}(c) shows that U-Net removes most of the external noise, but signal leakage still exists in the region contaminated by strong noise, as indicated by the blue arrows in Fig. \ref{fig:8}(c). The residual profile of the U-Net denoised result shows incomplete denoising for late arrivals, as indicated by the blue arrows in Fig. \ref{fig:8}(f). In contrast, SeisDeFM produces the cleanest denoised result and suppresses most external noise (Figs. \ref{fig:8}(d) and \ref{fig:8}(g)). In particular, as highlighted by the yellow circles in Figs. \ref{fig:8}(b)-\ref{fig:8}(d), both the Transformer and U-Net baselines exhibit severe signal leakage in the denoised results of late arrivals, which indicates that effective reflection energies are incorrectly removed together with the external noise. However, SeisDeFM preserves the continuity and amplitude of events for deep weak signals while removing noise effectively. These results demonstrate that SeisDeFM can better suppress external noise while preserving effective seismic signals.

\begin{figure}[htbp]
\centering
\includegraphics[width=\linewidth,height=0.6\textheight,keepaspectratio]{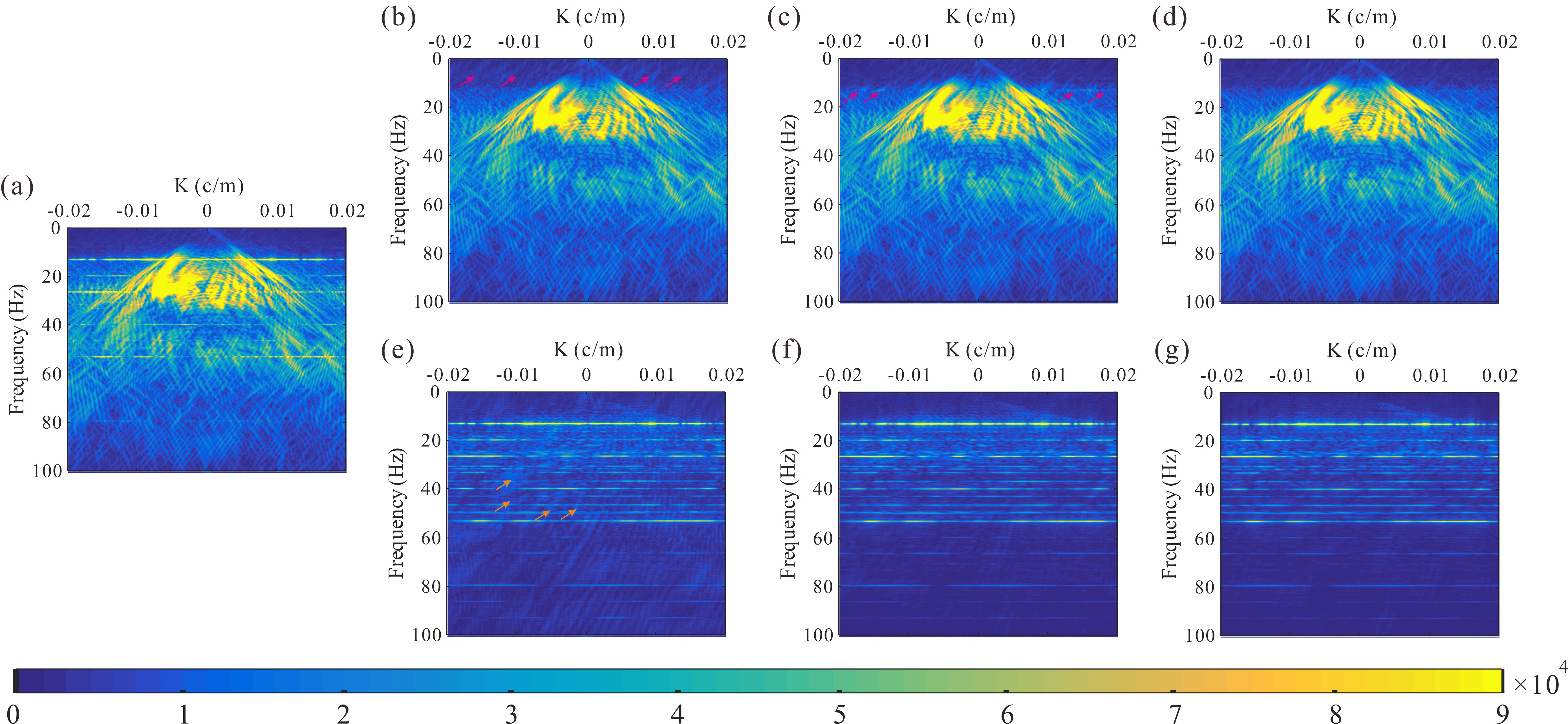}
\caption{F-K spectra comparisons of external-noise attenuation. (a)-(g) F-K spectra of the data shown in Figs. \ref{fig:8}(a)-\ref{fig:8}(g).}
\label{fig:9}
\end{figure}

To further evaluate the F-K-spectral behavior of different methods, we performed F-K domain analysis on the data shown in Fig.~\ref{fig:8}, with the results shown in Fig.~\ref{fig:9}. The denoised result obtained by Transformer introduces low-frequency noise that does not exist in the original noisy data, as indicated by the red arrows in Fig. \ref{fig:9}(b). Furthermore, the denoised residual of Transformer exhibits obvious signal energy leakage, as indicated by the yellow arrows in Fig. \ref{fig:9}(e). Compared with Transformer, the F-K spectrum of the U-Net denoised result shown in Fig. \ref{fig:9}(c) does not introduce extra low-frequency noise. Nevertheless, residual low-frequency external noise from the original noisy data remains, as indicated by the red arrows in Fig. \ref{fig:9}(c). In contrast, SeisDeFM yields the cleanest F-K spectrum for the denoised result, demonstrating its stronger ability to separate external noise from effective seismic signals.

\subsection{Surface Waves Attenuation}\label{sec:surface-wave-results}

Surface waves are coherent noise characterized by large amplitudes, low frequencies, and low apparent velocities. Surface waves often mask near-offset reflections and degrade subsequent seismic processing.

\begin{figure}[htbp]
\centering
\includegraphics[width=\linewidth,height=0.6\textheight,keepaspectratio]{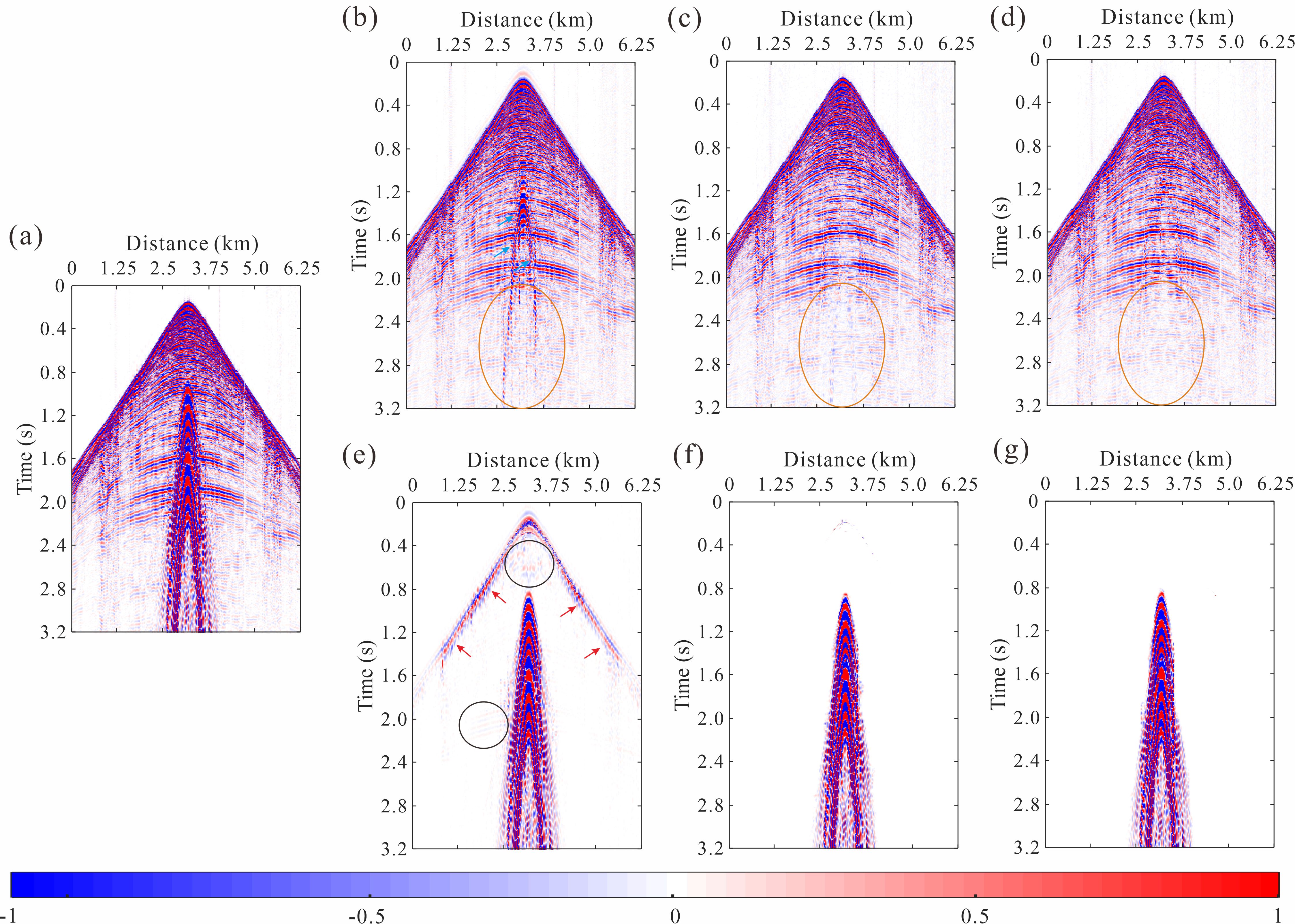}
\caption{Comparisons of surface-wave attenuation. (a) Raw noisy input data, (b)-(d) denoised results of Transformer, U-Net, and SeisDeFM, (e)-(g) the residual profiles between the input data shown in (a) and the denoised results shown in (b)-(d).}
\label{fig:10}
\end{figure}

The denoised results of the raw noisy data are shown in Fig.~\ref{fig:10}. Prominent surface-wave residuals can be observed in the denoised result of Transformer, as indicated by the blue arrows and yellow circle in Fig. \ref{fig:10}(b). Furthermore, the residual profile of the Transformer denoised result exhibits direct-wave damage (indicated by the red arrows in Fig. \ref{fig:10}(e)) and leakage of continuous reflections at early and later time (indicated by the black circles in Fig. \ref{fig:10}(e)). Although the denoised result of U-Net yields better surface-wave suppression compared with that of the Transformer, surface-wave residuals exist in regions with deep weak reflections, and the amplitude recovery of effective events is insufficient, as indicated by the yellow circle in Fig. \ref{fig:10}(c). In contrast, SeisDeFM suppresses surface waves more completely while preserving the continuity and amplitude fidelity of effective reflections (Figs. \ref{fig:10}(d) and \ref{fig:10}(g)), showing a better balance between noise attenuation and signal preservation. This test demonstrates that SeisDeFM achieves the best performance while requiring only~1,000 transfer-learning samples, indicating that the pre-trained seismic feature representations are also effective for suppressing coherent surface waves.

\begin{figure}[htbp]
\centering
\includegraphics[width=\linewidth,height=0.6\textheight,keepaspectratio]{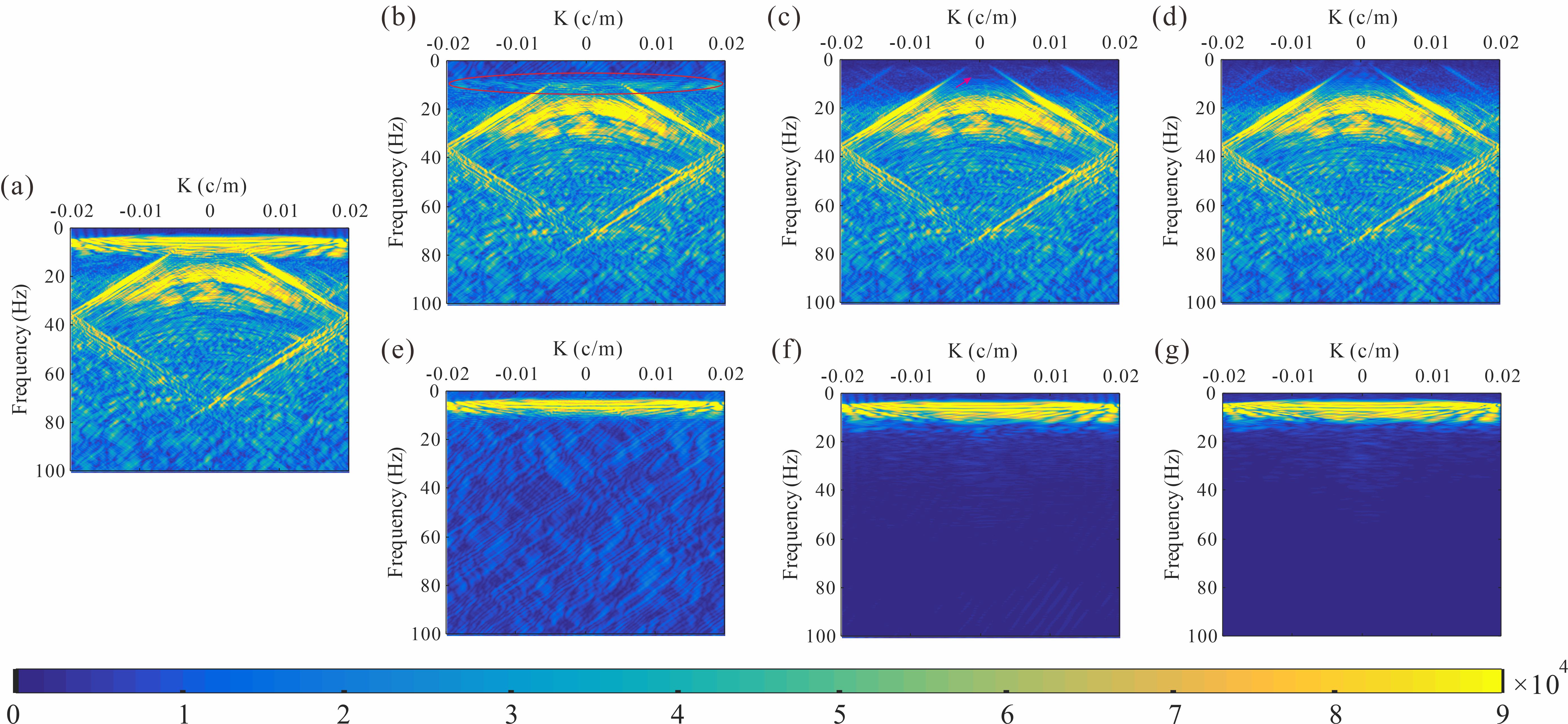}
\caption{F-K spectra comparisons of surface-wave attenuation. (a)-(g) F-K spectra of the data shown in Figs. \ref{fig:10}(a)-\ref{fig:10}(g).}
\label{fig:11}
\end{figure}

We further compared the corresponding F-K spectra for the data in Fig.~\ref{fig:10}, with the results presented in Fig.~\ref{fig:11}. The F-K spectrum of the Transformer's denoised result shows strong residual surface-wave energy in the low-frequency band, as indicated by the red circle in Fig. \ref{fig:11}(b). Moreover, in the F-K spectrum of the Transformer's residual profile shown in Fig. \ref{fig:11}(e), apart from the suppressed low-frequency surface-wave energy, the energy of leaked signals is widely distributed across the full frequency band. The F-K spectrum of U-Net's denoised result achieves considerable improvement over that of Transformer. The surface-wave energy in the low-frequency band is effectively eliminated, yet leakage of low-frequency signals remains, as indicated by the red arrow in Fig. \ref{fig:11}(c). In contrast, as shown in Figs. \ref{fig:11}(d) and \ref{fig:11}(g), SeisDeFM produces the F-K spectra of the denoised result and residual profile with weaker surface-wave energy and less signal leakage in the low-frequency band, demonstrating better performance for SeisDeFM in suppressing surface waves in the F-K domain.

\subsection{Random Noise Attenuation}\label{sec:random-noise-results}

We further performed a denoising test on field seismic data contaminated by random noise. Random noise does not follow a stable spatial pattern and may be distributed over a broad frequency range, making it difficult to attenuate without damaging weak reflection events.

\begin{figure}[htbp]
\centering
\includegraphics[width=\linewidth,height=0.6\textheight,keepaspectratio]{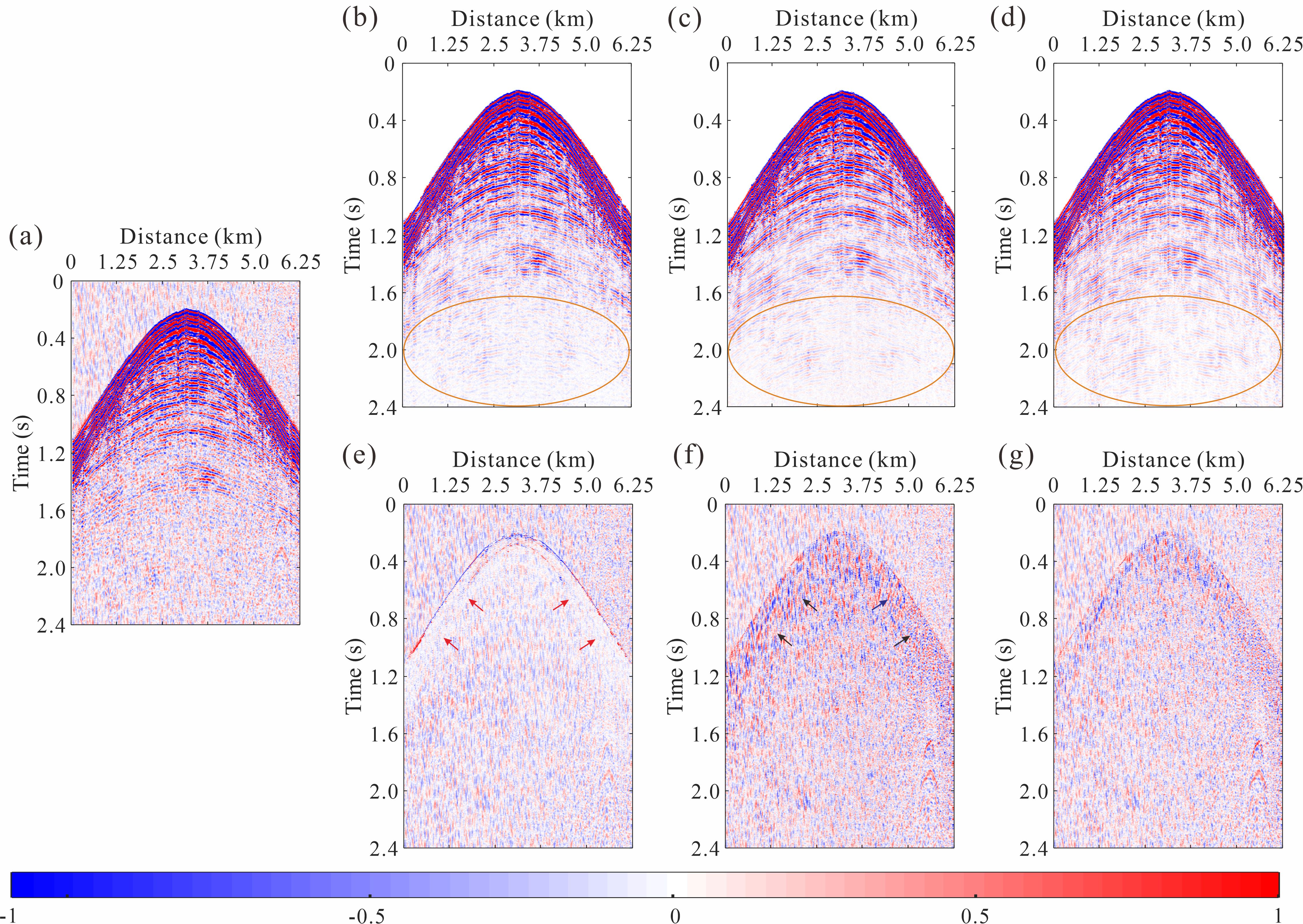}
\caption{Comparisons of random-noise attenuation. (a) Raw noisy input data, (b)-(d) denoised results of Transformer, U-Net, and SeisDeFM, (e)-(g) the residual profiles between the input data shown in (a) and the denoised results shown in (b)-(d).}
\label{fig:12}
\end{figure}

Comparison of denoised results for the field data under random-noise contamination is shown in Fig.~\ref{fig:12}. As indicated by the yellow circles in Figs. \ref{fig:12}(b)--\ref{fig:12}(d), the denoised results of the Transformer and U-Net baselines contain noise residuals in regions with deep weak signals and exhibit poor recovery of effective reflections. By contrast, SeisDeFM can effectively remove random noise while restoring the continuity and amplitude of deep reflections. As indicated by the red arrows in Fig. \ref{fig:12}(e), the absolute values at the region of early arrivals in the Transformer's denoising residual are small. This indicates that Transformer has poor denoising performance for the region where high-amplitude early arrivals overlap with random noise, making effective signal-noise separation difficult. As indicated by the black arrows in Fig. \ref{fig:12}(f), the absolute values at the region of early arrivals in U-Net's denoising residual are large. This indicates that U-Net misidentifies early arrivals as noise and attenuates them while removing random noise, resulting in signal leakage in this region. SeisDeFM performs the best in both noise suppression and signal preservation (Figs. \ref{fig:12}(d) and \ref{fig:12}(g)).

\begin{figure}[htbp]
\centering
\includegraphics[width=\linewidth,height=0.6\textheight,keepaspectratio]{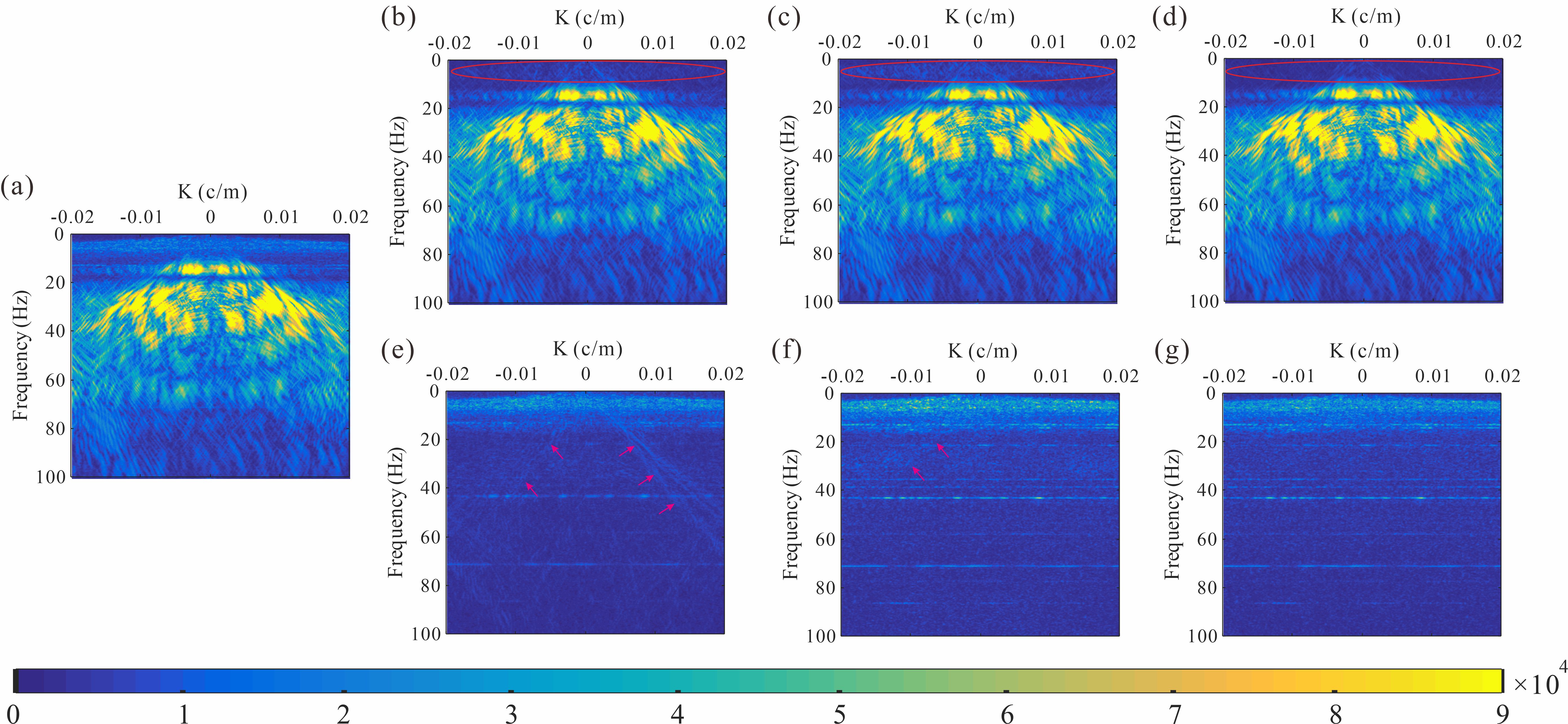}
\caption{F-K spectra comparisons of random noise attenuation. (a)-(g) F-K spectra of the data shown in Figs. \ref{fig:12}(a)-\ref{fig:12}(g).}
\label{fig:13}
\end{figure}

We further compared the corresponding F-K spectra for the data in Fig.~\ref{fig:12}, with the results presented in Fig.~\ref{fig:13}. As indicated by the red circles in Figs. \ref{fig:13}(b)--\ref{fig:13}(d), the residual energy of low-frequency noise exists in the F-K spectra of the denoised result from the Transformer and U-Net baselines. However, SeisDeFM can suppress low-frequency noise while preserving the effective signal energy in the low-frequency band. As indicated by the red arrows in Figs. \ref{fig:13}(e) and \ref{fig:13}(f), leaked signal energy exists in the F-K spectra of the denoising residuals from the Transformer and U-Net baselines. In contrast, SeisDeFM shows less random-noise energy in the denoised F-K spectrum (Fig. \ref{fig:13}(d)) and weaker signal energy in the residual F-K spectrum (Fig. \ref{fig:13}(g)). This test demonstrates that SeisDeFM not only suppresses random noise in the time-space domain but also better preserves the F-K spectral distribution of signals.

\section{Discussion}\label{sec:discussion}

\subsection{Ablation study of the pre-training strategy}\label{sec:ablation}

\begin{figure}[htbp]
\centering
\includegraphics[width=\linewidth,height=0.6\textheight,keepaspectratio]{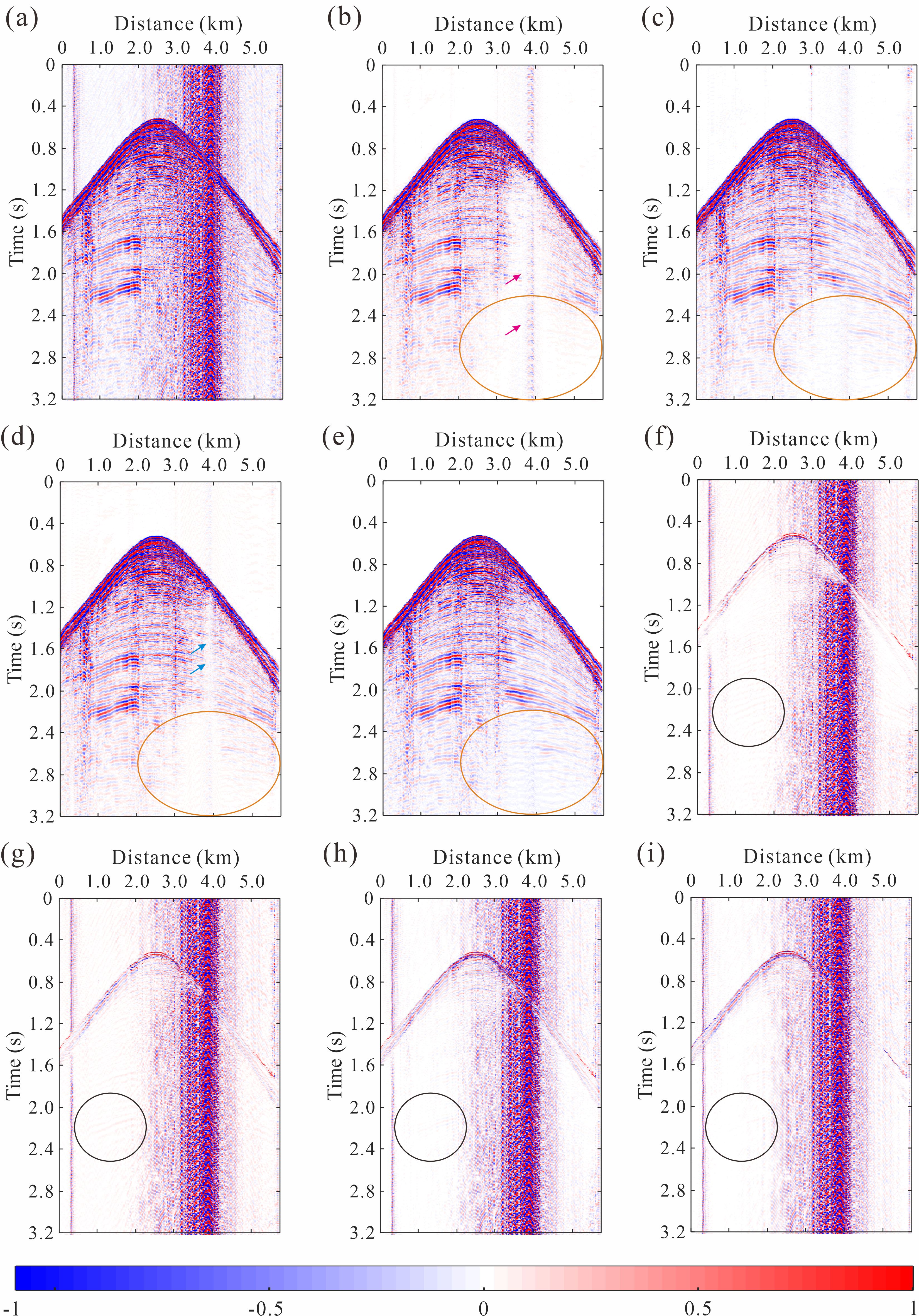}
\caption{Comparisons of external-noise attenuation via different pre-training strategies. (a) Raw noisy input data, (b)-(e) denoised results based on no pre-training, masked reconstruction, contrastive learning, and the joint strategy, (f)-(i) the residual profiles between the input data shown in (a) and the denoised results shown in (b)-(e).}
\label{fig:14}
\end{figure}

To analyze the contribution of different pre-training strategies, we conducted an ablation study on the external-noise attenuation task under an extremely low SNR condition. We compared four experimental settings: training from scratch without pre-training, pre-training with masked reconstruction only, pre-training with contrastive learning only, and the proposed joint pre-training strategy that combines masked reconstruction with contrastive learning. All models were trained via transfer learning for 30 epochs. The denoised result of a test sample is shown in Fig.~\ref{fig:14}. Without pre-training, obvious noise remains in the denoised result, as indicated by the red arrows in Fig. \ref{fig:14}(b). Moreover, severe signal leakage is exits in the deep reflections, as indicated by the yellow circle in Fig. \ref{fig:14}(b). As shown in Fig. \ref{fig:14}(c), the denoised result obtained under the mask-reconstruction-only pre-training strategy reduces external-noise residuals. This occurs because SeisDeFM acquires the capability to recover effective events during the self-supervised pre-training stage of mask reconstruction, enabling better separation of signals and noise after transfer learning for the downstream task. However, signal leakage still exists in deep weak reflection regions after denoising, as indicated by the yellow circle in Fig. \ref{fig:14}(c). As shown in Fig. \ref{fig:14}(d), the denoised result under the contrastive-learning-only pre-training strategy also reduces external-noise residuals. This is because SeisDeFM acquires the capability to distinguish positive and negative sample pairs during contrastive learning, which enables better separation of noise from noisy data. However, severe signal leakage appears in regions covered by strong noise (as indicated by the blue arrows in Fig. \ref{fig:14}(d)) and in deep weak reflection regions (as marked by the yellow circle in Fig. \ref{fig:14}(d)). In contrast, the joint pre-training strategy yields cleaner denoised results (Fig. \ref{fig:14}(e)) and mitigates signal leakage in the residual profiles (Fig. \ref{fig:14}(i)). As indicated by the black circles in Figs. \ref{fig:14}(f)--\ref{fig:14}(i), leakage of continuous events exists in the denoising residuals obtained from the no pre-training scheme, the mask-reconstruction-only pre-training strategy, and the contrastive-learning-only pre-training strategy. This ablation study demonstrates that context recovery and representation alignment in the pre-training stage are complementary for separating noises from seismic events. Although the joint pre-training strategy incurs higher training costs, the additional training time is little compared with the two standalone pre-training approaches, and it yields significant improvements in denoising performance.

\subsection{Comparisons of different transfer learning methods}\label{sec:transfer-comparison}

We also investigated the effects of different transfer learning methods on noise attenuation after pre-training, using test data containing random noise. Three transfer learning methods were compared: updating only the decoder of the SeisDeFM architecture, adding and updating lightweight adapter modules, and fine-tuning all model parameters. All methods were trained for 30 epochs with a learning rate of $ 1\times 10^{-4}$. In the decoder-only setting, the encoder was frozen and only the decoder and output layer were updated. In the adapter setting, lightweight trainable adapter modules were inserted into the original architecture, while the main encoder and decoder weights were frozen. The architecture of an adapter consists of normalization, dimension-reduction projection, activation, dropout, dimension-expansion projection, and residual addition. One adapter is embedded after each encoder layer and each decoder layer. In our case study of SeisDeFM, full-parameter fine-tuning was adopted, wherein all model parameters were fine-tuned so that pre-trained representations could fully adapt to the downstream denoising task.

Table~\ref{tab:2} lists the training cost and parameter statistics for different transfer learning methods. The adapter strategy has a slightly larger total parameter count due to the introduction of extra adapter modules, increasing the model size from 12.51~M to 12.64~M. However, only 142,562 parameters are trainable, corresponding to 1.13\% of the total parameters, and the training time is reduced to 80~min. The decoder-only method uses the same network structure as the full-parameter fine-tuning method, so the two strategies have identical total parameters and FLOPs. Because the encoder is frozen, decoder-only tuning updates 3.25~M parameters, accounting for 26.02\% of the total parameters, and requires 97~min for training. Full-parameter fine-tuning updates all 12.51~M parameters and takes 108~min for training. Although it introduces a slightly higher computational cost, it provides the network with greater flexibility to adapt to different levels of feature representations.


\begin{table}[htbp]
\centering
\caption{Quantitative comparisons of model complexity under different transfer learning methods.}
\label{tab:2}

\small
\setlength{\tabcolsep}{3.5pt}
\renewcommand{\arraystretch}{1.22}

\begin{tabularx}{\linewidth}{
@{}
>{\raggedright\arraybackslash}X
>{\centering\arraybackslash}m{0.12\linewidth}
>{\centering\arraybackslash}m{0.14\linewidth}
>{\centering\arraybackslash}m{0.16\linewidth}
>{\centering\arraybackslash}m{0.13\linewidth}
>{\centering\arraybackslash}m{0.11\linewidth}
@{}
}

\toprule

\multicolumn{1}{c}{\makecell[c]{\textbf{Method}}}
&
\makecell[c]{\textbf{Training}\\\textbf{Time}\\\textbf{(min)}}
&
\makecell[c]{\textbf{Total}\\\textbf{Parameters}}
&
\makecell[c]{\textbf{Trainable}\\\textbf{Parameters}}
&
\makecell[c]{\textbf{Trainable}\\\textbf{Ratio}}
&
\makecell[c]{\textbf{GFLOPs}}
\\

\midrule

Decoder-only
& 97
& 12.51 M
& 3,254,461
& 26.02\%
& 341.20
\\

Adapter
& 80
& 12.64 M
& 142,562
& 1.13\%
& 353.22
\\

\shortstack[l]{Full-parameter\\fine-tuning}
& 108
& 12.51 M
& 12,508,666
& 100.00\%
& 341.20
\\

\bottomrule

\end{tabularx}
\end{table}

\begin{figure}[htbp]
\centering
\includegraphics[width=\linewidth,height=0.6\textheight,keepaspectratio]{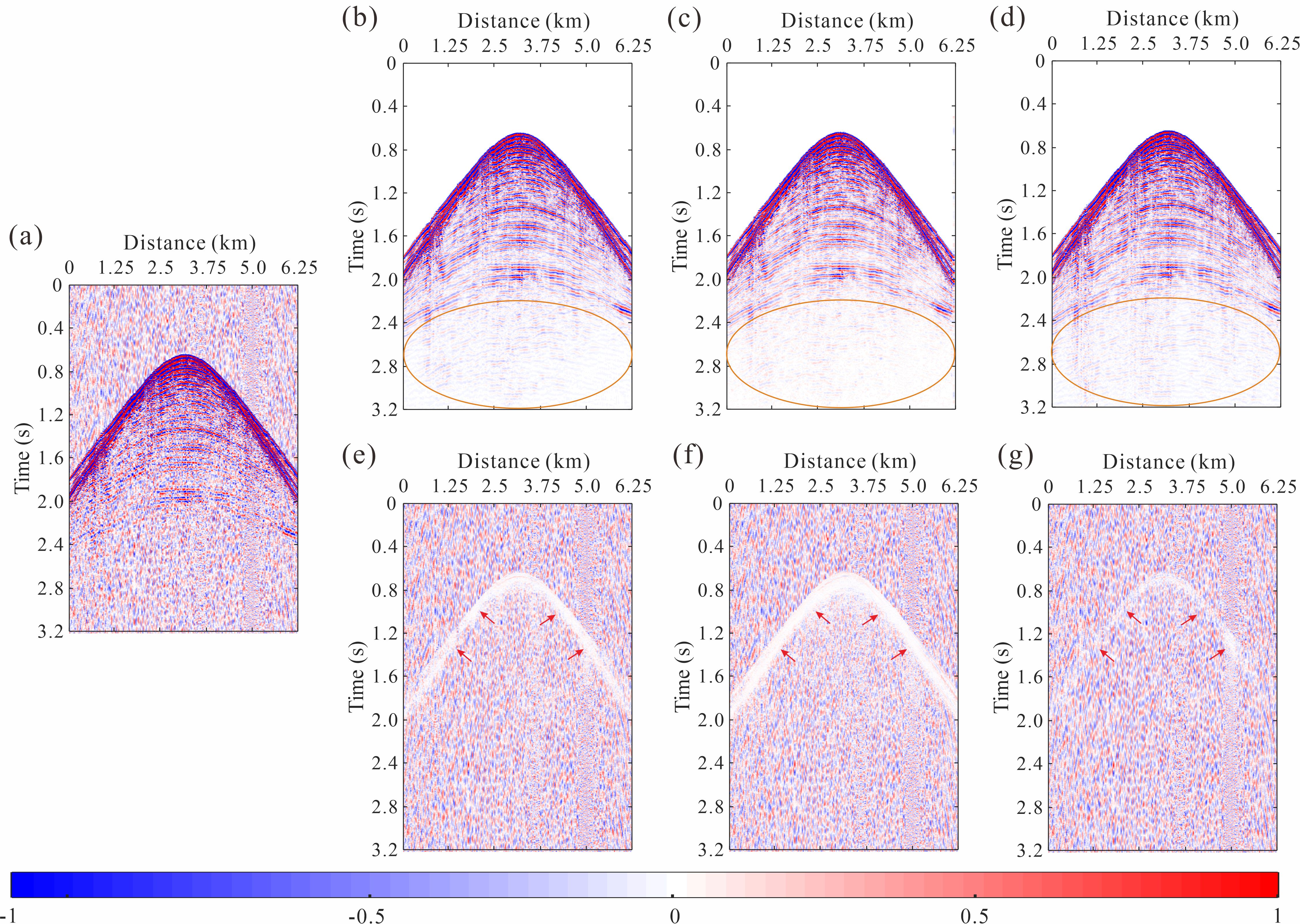}
\caption{Comparisons of random-noise attenuation via different transfer learning methods. (a) Raw noisy input data, (b)-(d) denoised results of decoder-only tuning, adapter-based tuning and full fine-tuning, (e)-(g) the residual profiles between the input data shown in (a) and the denoised results shown in (b)-(d).}
\label{fig:15}
\end{figure}

Comparisons of denoised results for the raw noisy data are presented in Fig.~\ref{fig:15}. The decoder-only and adapter-based tuning methods exhibit poor signal preservation in the deep regions contain weak reflections indicated by the yellow circles in Figs. \ref{fig:15}(b) and \ref{fig:15}(c). This indicates that partial-parameter updating may not be sufficient to correct the mismatch between the pre-trained representation and the target random-noise distribution. Furthermore, since the adapter-based method has the fewest trainable parameters, most signal leakage appears in its denoised result, as indicated by the yellow circles in Fig. \ref{fig:15}(c). As indicated by the red arrows in Figs. \ref{fig:15}(e)--\ref{fig:15}(f), the decoder-only tuning and adapter-based tuning fail to effectively separate signals from noise in early-arrival regions. This leads to small absolute values of the denoising residuals in these regions, and the adapter-based tuning strategy yields even worse performance. In contrast, full-parameter fine-tuning produces a cleaner denoised result and a residual profile with less signal leakage (Figs. \ref{fig:15}(d) and \ref{fig:15}(g)). This test demonstrates that updating the complete network is beneficial when the target noise pattern differs from the information learned during pre-training.

\FloatBarrier

\section{Conclusions}\label{sec:conclusions}

This paper presents a dedicated seismic denoising foundation model termed SeisDeFM to address the critical limitation of poor generalization among existing task-specific deep-learning denoising methods. SeisDeFM achieves stronger generalizability across diverse noise types and complex survey conditions. We present a case study of SeisDeFM, which adopts a hybrid architecture integrating a residual U-Net backbone for local multi-scale feature extraction and a trace-wise Transformer bottleneck for global wavefield dependency learning. It employs a joint self-supervised pre-training strategy combining masked wavefield reconstruction and contrastive learning, enabling the model to learn universal geophysical priors for seismic wavefields from large-scale clean data without paired noisy-clean ground-truth data. Extensive experiments on field seismic data contaminated by erratic noise, external noise, surface waves, and random noise demonstrate that SeisDeFM consistently outperforms the U-Net and Transformer baselines using only one-tenth of the labeled training data. This work validates that the foundation-model paradigm breaks the generalization ceiling of conventional task-specific denoising approaches, shifting seismic denoising from isolated task-specific mapping toward universal wavefield representation and efficient noise adaptation, and provides a promising pathway for large-scale industrial seismic data processing.

\section*{Acknowledgements}
This study was jointly supported by the National Natural Science Foundation of
China (Grant No. 42574169), the Deep Earth Probe and Mineral Resources Exploration—
National Science and Technology Major Project (Grant No. 2025ZD1007600),
and the Jilin Provincial Natural Science Foundation-Free Exploration Project (Grant
No. YDZJ202501ZYTS530).

\backmatter
\section*{Declarations}

\noindent\textbf{Conflict of interest.} The submission of this manuscript involves no conflicts of interest.

\nocite{*}
\bibliography{A_Review_of_Deep-learning-based_Seismic_Data_Denoising_and_Its_Promising_Paradigm_Shift_to_Foundation_Models}
\end{document}